\documentclass[sigconf]{acmart}
\usepackage{graphicx}
\usepackage{xr}
\usepackage{catchfilebetweentags}
\usepackage{multirow}
\usepackage{booktabs}
\usepackage{xspace}
\usepackage{enumitem}
\usepackage{subfig}
\usepackage{balance}
\usepackage{microtype}
\usepackage{ulem}
\usepackage{xcolor}
\usepackage[most]{tcolorbox}

\usepackage{framed}
\newenvironment{prompt}[1]{ \color{black}\MakeFramed {\advance\hsize-\width} {\FrameRestore} {\noindent#1}}{\endMakeFramed}

\begin{document}

\title{Prompt-Enhanced Software Vulnerability Detection Using ChatGPT}

\author{Chenyuan Zhang}
\orcid{0000-0002-2706-4237}
\affiliation{%
  \institution{Key Laboratory of Multimedia Trusted Perception and Efficient Computing, Ministry of Education of China, School of Informatics, Xiamen University}
  \country{China}
}
\email{zhangchenyuan@stu.xmu.edu.cn}
\authornote{The first two authors contributed equally.}
\author{Hao Liu}
\orcid{0000-0001-8087-1102}
\authornotemark[1]
\affiliation{%
  \department{Key Laboratory of Multimedia Trusted Perception and Efficient Computing, Ministry of Education of China, School of Informatics, Xiamen University}
  \country{China}
}
\email{haoliu@stu.xmu.edu.cn}

\author{Jiutian Zeng}
\author{Kejing Yang}
\author{Yuhong Li}
\affiliation{%
  \institution{Alibaba}
  \country{China}
}
\email{{zengjiutian.zjt, kejing.ykj, daniel.lyh}@alibaba-inc.com}

\author{Hui Li}
\orcid{0000-0001-9139-3855}
\affiliation{%
  \institution{Key Laboratory of Multimedia Trusted Perception and Efficient Computing, Ministry of Education of China, School of Informatics, Xiamen University}
  \country{China}
}
\email{hui@xmu.edu.cn}

\begin{abstract}

With the increase in software vulnerabilities that cause significant economic and social losses, automatic vulnerability detection has become essential in software development and maintenance. Recently, large language models (LLMs) like GPT have received considerable attention due to their stunning intelligence, and some studies consider using ChatGPT for vulnerability detection. However, they do not fully consider the characteristics of LLMs, since their designed questions to ChatGPT are simple without a specific prompt design tailored for vulnerability detection. This paper launches a study on the performance of software vulnerability detection using ChatGPT with different prompt designs. Firstly, we complement previous work by applying various improvements to the basic prompt. Moreover, we incorporate structural and sequential auxiliary information to improve the prompt design. Besides, we leverage ChatGPT's ability of memorizing multi-round dialogue to design suitable prompts for vulnerability detection. We conduct extensive experiments on two vulnerability datasets to demonstrate the effectiveness of prompt-enhanced vulnerability detection using ChatGPT. We also analyze the merit and demerit of using ChatGPT for vulnerability detection. Repository: \url{https://github.com/KDEGroup/LLMVulnerabilityDetection}.
\end{abstract}

\keywords{software vulnerability detection, prompt, large language model, chatgpt}

\maketitle

\section{Introduction}
\label{sec:intro}

Software has become an indispensable part of our digital society.
However, an increasing number of software vulnerabilities are causing significant economic and social losses~\cite{NguyenNNLTP22}.
Some studies have shown that security vulnerabilities exist in a significant number of open-source code repositories, and nearly half of them contain high-risk vulnerabilities~\cite{LiZXJZC22, LippBP22}, suggesting that software vulnerability detection needs further improvement and it remains an unsolved and challenging problem in the software industry~\cite{LinWHZX20}. 
Consequently, it is imperative to implement intelligent automatic software vulnerability detection methods to protect software security better.
 
Traditional vulnerability detection approaches are mainly based on rules and classical machine learning (ML) techniques. 
Rule-based approaches~\cite{Checkmarx,SuiX16,GaoMSSZMMDDZC18} use pre-defined vulnerability rules for detection. 
These user-defined features highly rely on expert knowledge and are generally labor-intensive, making it difficult to deploy them to cover different software vulnerabilities. 
ML-based vulnerability detection approaches~\cite{SharBT15, GriecoGURFM16, ScandariatoWHJ14, PangXN15} learn latent features of the vulnerable code snippets, providing better generalization and detection accuracy compared to  rule-based approaches.
They usually extract code features by leveraging Word2vec~\cite{abs-1301-3781} or continuous bag-of-words (CBOW)~\cite{abs-1301-3781} and then applying classical machine learning algorithms like Support Vector Machine and Logistic Regression
However, they rely on coarse-grained patterns of vulnerabilities due to their shallow architectures and they cannot achieve accurate vulnerability detection.

Inspired by the success of deep learning (DL), several attempts~\cite{DiwanLF22, CaoSBWLT22,CuiHJFY21,LinZLPVMX21} aim to detect vulnerabilities using deep neural networks. 
Compared to traditional techniques, they can capture implicit and complicated vulnerability patterns from source code better. 
In the meanwhile, various sophisticated code modeling methods, such as control flow graph (CFG), program dependence graph (PDG) and data flow graph (DFG), are leveraged by DL-based methods to enrich code semantics~\cite{LiZXJZC22, SharT13, LinZLPX17, CuiHJFY21}. 
There are also some studies adopting texts written in natural language (e.g., code summaries, instructions, and commit messages) to enhance representation learning for DL-based vulnerability detection~\cite{LeNLPMVQ19, LeNLPMVQ19, ChoiJOC17, abs-1808-09897, PengMLLZJ15,ZhouLSD019, WangYTTHFFBW21, MonKCM23}. 
Nevertheless, most existing DL-based approaches are not general detection methods, as they are programming-language-specific or vulnerable-type-specific.

In recent years, large-scale language models (LLMs) like GPT have received increasing attention due to their stunning intelligence~\cite{abs-2303-18223}. 
LLMs are trained on large corpora using powerful computation resources.
They typically contain billions of model parameters, allowing them to capture complex patterns and reveal general intelligence in many tasks that were thought to be difficult for AI to complete in the near future successfully. 
A representative example is ChatGPT\footnote{\url{https://chat.openai.com}}, a versatile chatbot empowered by the LLM GPT that allows users to have human-like conversations to receive desirable answers, including but not limited to composition (e.g., music, teleplays, fairy tales, and student essays), programming and answering test questions.  
As a result, some recent works have considered using ChatGPT to improve software vulnerability detection~\cite{abs-2301-08653,abs-2304-08191}. 
Sobania et al.~\cite{abs-2301-08653} evaluate ChatGPT’s bug-fixing performance on a standard benchmark. 
Cao et al.~\cite{abs-2304-08191} explore ChatGPT’s ability to fix DL programs. 
However, these methods do not fully consider the characteristics of LLMs, since their designed questions to ChatGPT are simple without a specific prompt design tailored for vulnerability detection using LLMs.
Prompt engineering modifies the original user input using a textual prompt, and the resulting text has some unfilled slots.
Then, LLMs are asked to fill the unfilled information which is actually the answer that the user requires. 
Prompt engineering has been shown to be promising in exploring the potential of LLMs~\cite{LiuYFJHN23}, although it has not been fully considered in software vulnerability detection.
Besides, unlike traditional ML or DL methods, LLMs can solve complicated tasks through multiple reasoning steps (i.e., chain-of-thought), which is ignored by existing works on vulnerability detection.

To address the above issues, this paper launches a study on the performance of vulnerability detection using ChatGPT 4 with different prompt designs.
ChatGPT 4 is the newest version of ChatGPT when we conduct this study, and the term ChatGPT refers to ChatGPT 4 in this paper.
The contributions of this work are summarised below:
\begin{itemize}[leftmargin=12pt,topsep=1pt,itemsep=0.3pt]

\item We complement previous work by applying various improvements to the basic prompt and investigating the vulnerability detection capabilities of ChatGPT on our collected vulnerability datasets covering two programming languages.

\item We incorporate structural and sequential auxiliary information of the source code in prompt designs, which is shown to be helpful for ChatGPT to detect vulnerabilities better. To our best knowledge, this is the first time that traditional structural and sequential code modeling methods can be directly used in ChatGPT-based vulnerability detection.

\item We leverage ChatGPT's ability of memorizing multi-round dialogue in vulnerability detection. Through our designed chain-of-thought prompting, the performance of ChatGPT on vulnerability detection gets further improvements. This is also not considered in previous works.

\item We conduct extensive experiments on two vulnerability datasets to demonstrate the effectiveness of prompt-enhanced vulnerability detection using ChatGPT. We also analyze the merit and demerit of using ChatGPT for vulnerability detection.

\end{itemize}

The remainder of this paper is organized as follows:
Sec.~\ref{sec:relatedwork} presents the related work of this paper.
Sec.~\ref{sec:data} describes the data used in our study, and Sec.~\ref{sec:prompt} demonstrates our prompt designs for enhancing software vulnerability detection using ChatGPT.
Sec.~\ref{sec:res} provides the experimental results and discussions on the vulnerability detection capability of ChatGPT.
Sec.~\ref{sec:threats} describes the potential threats of validity for this work.
Sec.~\ref{sec:con} concludes this work.


\section{Related Work}
\label{sec:relatedwork}

\subsection{Large language models}

Large language models (LLMs) have recently revolutionized AI, exhibiting remarkable prowess across a wide range of tasks~\cite{BrownMRSKDNSSAA20,abs-2206-04615,abs-2108-07258}. 
LLMs are based on the idea of language modeling (LM).
LM is a prevalent approach to model the generative likelihood of word sequences so as to predict the future tokens.
Unlike traditional LM approaches, LLMs are trained on a huge volume of data using powerful computation resources.
The versatility and adaptability of LLMs are deemed to be credited with their billion-scale parameters~\cite{abs-2303-18223} that have never been achieved in the past.

Representative LLMs include but not limited to OpenAI’s GPT~\cite{BrownMRSKDNSSAA20,abs-2303-08774}, Google’s PaLM and Bard~\cite{abs-2204-02311,abs-2305-10403} and DeepMind’s Chinchilla~\cite{abs-2203-15556}.
However, some of them can be accessed through provided APIs while others are not accessible~\cite{abs-2306-08568}. 
There are many other LLMs that are open-source.
EleutherAI has contributed GPT-NeoX-20B~\cite{abs-2204-06745} and GPT-J-6B\footnote{\url{https://github.com/kingoflolz/mesh-transformer-jax}}. Google has released UL2-20B~\cite{Tay00GW0CBSZZHM23}. Tsinghua University has introduced GLM-130B~\cite{ZengLDWL0YXZXTM23}. 
The Transformer architecture-based language models OPT~\cite{abs-2205-01068} and LLaMA~\cite{abs-2302-13971} released by Meta have also attracted attention recently.

\subsection{Prompt Engineering}

Prompting-based learning has become a prevalent learning paradigm for LM.
Instead of using pre-trained language models for downstream tasks via objective engineering, prompting-based learning reformulates the downstream tasks through a textual prompt so that they are close to tasks solved during the original LM training~\cite{LiuYFJHN23}.
Well-structured prompts have been shown to be promising in improving the performance of LLMs in various downstream tasks~\cite{ZhaoWFK021,abs-2304-05970}. 
Therefore, various prompt design paradigms have proliferated~\cite{LiuYFJHN23}.  

Regarding the format of prompts, some works explore prompt search for appropriately discrete prompt~\cite{GaoFC20,JiangXAN20,LiuSZDCC22,ShinRLWS20}. 
Meanwhile, some works utilize continuous vector (i.e., embeddings) as prompts~\cite{GuHLH22,LesterAC21,LiL20,abs-2103-10385,QinE21,ZhouYLL22}. 

Some work has examined the effect of prompts on generative models. 
For example, Liu et al.~\cite{LiuC22a} investigate how different prompt keywords affect image generation. Maddigan and Susnjak have explored using prompts to help LLMs generate visualizations~\cite{abs-2302-02094}.  
Liu et al.~\cite{abs-2305-08360} propose different prompt designs for two code generation tasks. 
There are a few recent works studying exploring the potential of ChatGPT on the software vulnerability detection task.
Cao et al.~\cite{abs-2304-08191} design enhanced prompt templates to apply ChatGPT in Deep Learning program repair.
White et al.~\cite{abs-2303-07839} have explored several prompt patterns that can be applied to improve requirements elicitation, rapid prototyping, code quality, deployment, and testing.
However, as discussed in Sec.~\ref{sec:intro}, existing works do not fully consider the characteristics of LLMs when detecting vulnerabilities.

\subsection{Software Vulnerability Detection}

\subsubsection{Traditional Vulnerability Detection}

Traditional vulnerability detection methods rely on feature engineering and traditional machine learning techniques.
They generally learn the code patterns from the handcrafted features extracted from static and/or dynamic code analysis. 
Some early studies~\cite{EnglerCC01, ZhangXWZXG13, LiZ05, WenHCXZJ15} apply rules drawn from experience as ``templates'' to detect potential vulnerabilities. 
Shar et al.~\cite{SharBT15} combine the hybrid program analysis with supervised classifiers like logistic regression (LR) and random forest (RF) for vulnerability detection.
Wang et al.~\cite{WangLT16} firstly utilize ASTs for detecting vulnerabilities in Java programs. 
They combine function-level ASTs and semantic representations generated by deep belief network at file-level to train traditional machine learning models like logistic regression.
Grieco et al.~\cite{GriecoGURFM16} present VDiscover, which integrates both static and dynamical features using LR, RF and multi-layer perception (MLP) on a large-scale Debian program dataset. 
Harer et al.~\cite{abs-1803-04497} propose a data-driven approach specifically applied to C/C++ programs. 
They leverage features from both the source code and the build process, and use deep neural networks (DNNs) and conventional models like random forests for vulnerability detection.
Scandariato et al.~\cite{ScandariatoWHJ14} use the bag-of-words representation in which a software component is seen as a series of terms with associated frequencies as features used in vulnerability detection.
Peng et al.~\cite{PangXN15} combine N-gram analysis and feature selection algorithms for vulnerability detection and features are defined as continuous sequences of token in source code files.

\subsubsection{Deep Learning Based Vulnerability Detection}
\label{sec:DL}

The success of deep learning techniques (DL) in various fields has inspired researchers to apply DL in the vulnerability detection task~\cite{ChakrabortyKDR22}.
Compared to traditional methods, DL based methods can automatically capture vulnerability features, reducing the cost of feature engineering.

Existing DL-based methods can be classified into two categories: sequence-based methods~\cite{LiZXO0WDZ18, ZouWXLJ21, RussellKHLHOEM18,ChengZ0S22}
and graph-based methods~\cite{SharT13, LinZLPX17, LinZLPXVM18, abs-1708-02368, LiZXJZC22, DongWLXZ18, LinZLPVMX21, CuiHJFY21, HinKCB22, DiwanLF22, CaoSBWLT22}. 

Sequence-based methods adopt DNNs to model sequential code entities.
Li et al.~\cite{LiZXO0WDZ18} propose the ``code gadget'' definition to represent the program from the perspective of slice. 
They obtain code gadgets by extracting and assembling the library/API function calls and the corresponding slices in each program. 
Then a Bi-LSTM network is applied to detect whether the code is vulnerable based on code gadgets features. 
Zou et al.~\cite{ZouWXLJ21} regard ``code gadget'' as the global features and extend the idea of Li et al.~\cite{LiZXO0WDZ18} by introducing the so-called ``code attention'' as the ``localized'' features.
They leverage Bi-LSTM to fuse both features and conduct classification.
Cheng et al.~\cite{ChengZ0S22} present ContraFlow, a path-sensitive code embedding approach using a pre-trained value-flow path encoder.
They design a value-flow path selection algorithm and apply feasibility analysis to filter valued sequential paths as the input.
There are also works~\cite{RussellKHLHOEM18} applying CNNs to encode sequential code data for function-level vulnerability detection. 

Graph-based methods leverage DL methods to model graph data of programs, including abstract syntax tree (AST), data flow graph (DFG), control flow graph (CFG), project dependency graph (PDG), in vulnerability detection.
Shar et al.~\cite{SharT13} propose to use various static code attributes extracted from CFGs to detect SQL injection and cross-site scripting vulnerabilities in open source PHP projects. 
Lin et al.~\cite{LinZLPX17} utilize Bi-LSTM to process ASTs at function-level on three open source projects from Github for vulnerability detection.
Another AST based approach proposed in Dam et al.~\cite{abs-1708-02368} leverages LSTM for file-level vulnerability detection in open-source Android projects. 
Li et al.~\cite{LiZXJZC22} 
utilize various types of RNNs (Bi-LSTM and Bi-GRU) to 
model CFGs, DFGs and PDGs in order to detect software vulnerabilities.
Cui et al.~\cite{CuiHJFY21} propose VulDetector, a static analysis tool where the key is a weighted feature graph (WFG) model to slice the CFGs, for vulnerability detection. 
Hin et al.~\cite{HinKCB22} present LineVD, which uses a transformer-based model for encoding and Graph Neural Networks for processing control and data dependencies among statements.

\section{Data}
\label{sec:data}

This section describes the data used in our study.

\subsection{Data Collection}
\label{sec:Javadata}
We adopt two datasets containing functions with or without vulnerabilities for evaluating vulnerability detection.
One dataset contains Java functions and the other dataset contains C/C++ functions.

For the Java dataset, we collect vulnerable code samples from the Software Assurance Reference Dataset (SARD)\footnote{\url{https://samate.nist.gov/SARD}}. 
SARD is a standard vulnerable database where the data is derived from the Software Assurance Metrics And Tool Evaluation (SAMATE) project of the National Institute of Standards and Technology (NIST). Each program in the SARD contains both the real-world vulnerable samples and synthetic test cases, and is accompanied by a label as good (i.e., non-vulnerable code), bad (i.e, vulnerable code) or mixed (i.e., containing both vulnerable code and corresponding patched version) with a unique CWE ID. We obtained all Java vulnerable data published before June 8th, 2023 covering 46,415 projects. 

For the C/C++ dataset, we use the recent research release benchmark ~\cite{LiZXJZC22}, which was collected from the National Vulnerability Database (NVD)\footnote{\url{https://nvd.nist.gov/}}. NVD contains vulnerabilities in software products (i.e., software systems) and possibly diff files describing the difference between a vulnerable piece of code and its patched version. We focus on  the .c or .cpp files that contain some vulnerability (corresponding to a CVE ID) or its patched version.

\begin{figure*}[t]
    \centering
    \includegraphics[width=1\linewidth]{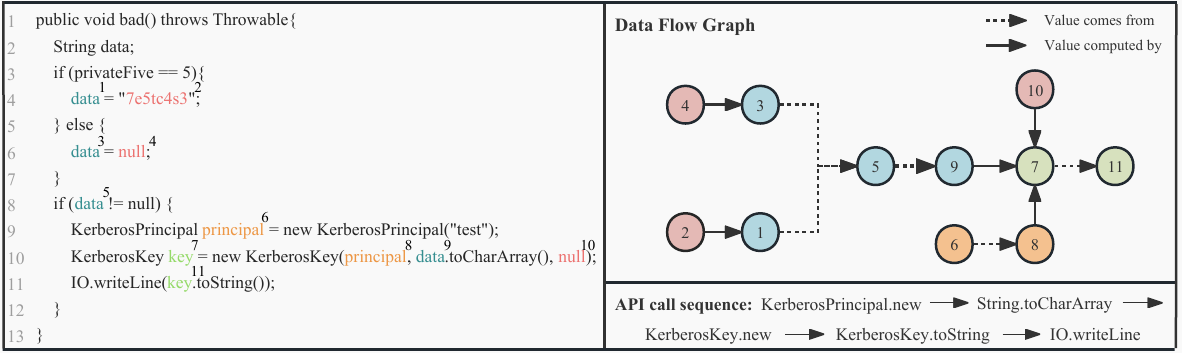}
    \caption{An example of a Java function and its auxiliary information.}
    \label{fig:apiseq}
\end{figure*}

\subsection{Data Pre-processing}
\label{sec:preprocess}

We perform the following data pre-processing steps on the two datasets:
\begin{enumerate}[leftmargin=14pt,topsep=1pt,itemsep=0.3pt]

\item We removed duplicated code and all annotations in each function because they explicitly indicate the vulnerability. 
In addition, we replace ``bad'' , ``good'' ,``VULN" and ``PATCHED" in function names with ``func''.

\item Considering the limited input length of ChatGPT, we set the maximum length of input to be 700 to ensure that the input sequence after concatenating additional information (i.e., prompt) is not truncated by the system. 

\item We filter out functions with less than three lines as they possibly do not have sufficient information for vulnerability detection.

\item For Java data, as the class name in dataset contains vulnerability information (i.e., vulnerability type) and we only focus on function-level vulnerability detection, we filter out those samples that call functions in other Java classes. For C/C++, we removed CVE IDs in function names that have obvious vulnerability hints.

\end{enumerate}

For Java data, a ``bad'' function can be associated with multiple ``good'' ones (i.e., patch versions) in ``mixed'' cases. We randomly selected an eligible ``good' one for such cases to balance the number of vulnerable and non-vulnerable samples. 
By examining the data distribution of vulnerabilities in the Java dataset, we find that more than 80\% of the samples belong to top-50 largest vulnerability types. 
Therefore, we consider samples in these 50 vulnerability types since we also investigate the effectiveness of ChatGPT on detecting different vulnerability types and the evaluation results on vulnerability types with few samples may not be precise.
Finally, as shown in Tab.~\ref{tab:data}, we get 1,171 vulnerable functions and 917 non-vulnerable functions for the Java dataset, and the C/C++ dataset contains 1,015 vulnerable functions and 922 non-vulnerable functions.

\begin{table}[t]
\caption{The statistics of the data.}
\label{tab:data}
\begin{tabular}{|c|c|c|c|c|}
\hline
\textbf{Language}  & \textbf{Types} & \textbf{\#Vul} & \textbf{\#Non-Vul} & \textbf{All} \\ \hline
Java  & 50    & 1,171       & 917        & 2,088         \\ \hline
C/C++ & 39    & 1,015      & 922 & 1,937
\\ \hline
\end{tabular}
\end{table}

\subsection{Feature Extraction}
\label{sec:feature}

In our prompt design, two types of auxiliary information, API call sequences and data flow, are utilized.
In this section, we illustrate how we extract them from the two datasets.

\subsubsection{API Call Sequence Extraction}
\label{sec:api}
To extract the API call sequences, we first leverage the Python library \textsc{tree-sitter}\footnote{\url{https://tree-sitter.github.io/tree-sitter}} to parse code files into Abstract Syntax Trees (ASTs). 
For C/C++, we extract all the nodes whose type is "call\_expression" and add all the names by order.
For Java, we follow the same steps as described in DEEPAPI~\cite{GuZZK16} to traverse each AST. The API call sequence of each Java function can be extracted as follows:
\begin{itemize}[leftmargin=12pt,topsep=1pt,itemsep=0.3pt]
\item For each constructor invocation \textit{new C()}, we add \textit{C.new} to the API call sequence.
\item For each function call \textit{o.m()} where \textit{o} is an instance of the class \textit{C}, we add \textit{C.m} to the API call sequence.
\item For a function call passed as a parameter, we add the function before the calling function. For example, for \textit{$o_{1}.f_{1}(o_{2}.f_{2}(),$}
\textit{$ o_{3}.f_{3}())$}, we add the API call sequence \textit{$C_{2}.f_{2}$}-\textit{$C_{3}.f_{3}$}-\textit{$C_{1}.f_{1}$}, where \textit{$C_{i}$} is the class of the instance \textit{$o_{i}$}.
\item For a sequence of statements \textit{$s_{1};s_{2};...;s_{n}$}, we extract the API call sequence \textit{$a_{i}$} from each statement \textit{$s_{i}$}, and concatenate them together in order to construct the API call sequence \textit{$a_{1}$}\textit{$a_{2}$}--...-\textit{$a_{n}$}.
\item For conditional statements such as \textit{$if (s_{1}) \{s_{2};\} else \{s_{3};\}$}, we create a sequence from all possible branches, that is, \textit{$a_{1}$}-\textit{$a_{2}$}-\textit{$a_{3}$}, where \textit{$a_i$} is the API call sequence extracted from the statement \textit{$s_{i}$}.
\item For loop statements such as \textit{$while (s_{1})\{s_{2};\}$}, we add the sequence \textit{$a_{1}$}-\textit{$a_{2}$}, where \textit{$a_{1}$} and \textit{$a_{2}$} are the API sequences extracted from the statement \textit{$s_{1}$} and \textit{$s_2$}, respectively.
\end{itemize}
The lower right part of Fig.~\ref{fig:apiseq} provides an example of the extracted API call sequence from a Java method. 
Note that, in our study, the function name ``bad'' will be replaced with ``func'' as described in Sec.~\ref{sec:preprocess}.

\subsubsection{Data Flow Graph Extraction}
\label{sec:dfg}
Data flow represents the relation of ”where-the-value-comes-from” between variables. Given the source code $C$, we refer to GraphCodeBERT's~\cite{GuoRLFT0ZDSFTDC21} extraction method and perform the following steps to extract the data flow:
\begin{itemize}[leftmargin=12pt,topsep=1pt,itemsep=0.3pt]
    \item Parse $C$ into an Abstract Syntax Tree (AST) using tree-sitter. 
    \item Identify the variables by using the syntactic information and terminals (leaves) of the AST, which are denoted as $V = \{v_1, v_2, ... , v_k\}$.
    \item Consider the variables as nodes of the graph, and for each variable in $V$, add the directed edge $\varepsilon =<v_i, v_j>$ to the dictionary $D$ as a tuple $(v_i, p_i, v_j, p_j)$, where $p_i$ and $p_j$ denote the position of $v_i$ and $v_j$. The tuple indicates that $v_i$ comes from or is computed by $v_j$.
\end{itemize}
The upper right part of Fig.~\ref{fig:apiseq} shows the extracted data flow graph of a Java method. 

\section{Prompt Design}
\label{sec:prompt}

This section introduces our designed prompts for enhancing ChatGPT on software vulnerability detection. 
For simplicity, we use \textbf{P$_{x}$} to represent each prompt, where $x$ denotes the compositions of the prompt, and we will explain each $x$ when we explain the corresponding prompt design for the first time. 

\subsection{Basic Prompting}
\label{sec:basic prompt}

Firstly, to conduct vulnerability detection via ChatGPT, it is essential to have a \underline{b}asic prompt (\textbf{P}$_{\text{b}}$). 
We use the following basic prompt in this study, and we ask ChatGPT to give an explicit answer (Yes or No) to clarify the result.
\begin{prompt}{\textbf{P}$_{\text{b}}$:}
Is the following program buggy? Please answer Yes or No. [CODE]
\end{prompt}
\noindent where [CODE] indicates the test function.
 
Following OpenAI's gpt-best-practices document~\cite{GBP}, we further propose the \underline{r}ole-based \underline{b}asic prompt (\textbf{P}$_{\text{r-b}}$) to remind ChatGPT of its job (i.e., detect vulnerabilities) so that it focuses on the vulnerability issue:
\begin{prompt}{\textbf{P}$_{\text{r-b}}$:}
I want you to act as a vulnerability detection system. My first request is ``Is the following program buggy?'' Please answer Yes or No. [CODE]
\end{prompt}

When the prompt contains some key tokens (e.g., the keyword ``buggy'' in the prompt), it is possible that ChatGPT guesses the answer simply based on the keywords instead of analyzing the test function, i.e., merely answering ``No'' due to the keyword ``buggy''.
To verify whether ChatGPT will make a biased judgment due to some key tokens in the prompt,
we designed a \textit{\underline{r}everse-question} prompt to enhance the \underline{r}ole-based \underline{b}asic prompt (\textbf{P}$_{\text{r-r-b}}$):
\begin{prompt}{\textbf{P}$_{\text{r-r-b}}$:}
I want you to act as a vulnerability detection system. My first request is ``Is the following program correct?'' Please answer Yes or No. [CODE]
\end{prompt}

\subsection{Prompting with Auxiliary Information}

Auxiliary information, such as CFG, DFG, PDG and API call sequences, is prevalently used by DL-based vulnerability detection methods~\cite{LiZXJZC22, SharT13, LinZLPX17,CuiHJFY21}.
This inspires us to consider whether we can utilize them to improve the prompt and achieve better vulnerability detection.

\subsubsection{The Necessity of Auxiliary Information}

Before incorporating auxiliary information into prompts, we first examine \textit{what types of auxiliary information are beneficial to detecting vulnerabilities using ChatGPT.}
For each dataset, we randomly sample 20 functions and ask ChatGPT the following question:
\begin{prompt}
Please output [data flow/control flow/program dependence] of the following code: [CODE]
\end{prompt}

As an example, the response of ChatGPT w.r.t. the code shown in Fig.~\ref{fig:apiseq} is shown below:
\begin{prompt}
\noindent\textbf{The answer for data flow}:\newline
1. the data value of data at 5th token comes from privateFive at 1st token \newline
2. the data value of data at 5th token comes from the constant ``7e5tc4s3'' at 5th token \newline
3. the data value of data at 8th token comes from privateFive at 1st token \newline
4. the data value of data at 8th token comes from the constant null at 8th token \newline
5. the data value of principal at 15th token comes from the constant ``test'' at 15th token \newline
6. the data value of key at 20th token comes from principal at 15th token \newline
7. the data value of key at 20th token comes from data at 5th token or 8th token \newline
8. the data value of key at 20th token comes from the constant null at 20th token 
\end{prompt}
For the response w.r.t. data flow, we can observe that it does not recognize two possible paths for the variable ``data'' in the second if statement. 
Besides, in the third line, the variable ``data'' does not come from the variable ``privateFive''. The incorrect response w.r.t. data flow shows that ChatGPT cannot fully understand the data flow of a function solely based on the source code. 

The answer of ChatGPT w.r.t. control flow is shown as follows:
\begin{prompt}
\noindent\textbf{The answer for control flow}:\newline
1. Check if privateFive equals 5.\newline
2. If privateFive is 5, set data to ``7e5tc4s3''. Otherwise, set data to null.\newline
3. Check if data is not null. If it's not null, proceed to step 4. Otherwise, exit the function.\newline
4. Create a new KerberosPrincipal object called principal with the argument ``test''.\newline
5. Create a new KerberosKey object called key with the arguments principal, data.toCharArray(), and null.\newline
6. Output the key as a string using IO.writeLine().\newline
7. End of function.
\end{prompt}
For control flow, we can observe that ChatGPT can accurately identify the execution path of each statement, indicating that it can understand the function's control flow. 

The answer of ChatGPT w.r.t. program dependence is shown below:
\begin{prompt}
\noindent\textbf{The answer for program dependence}:\newline
1. Control Dependence:\newline
- `if (privateFive == 5)' statement is control dependent on the function entry (bad() method). \newline
- `if (data != null)' statement is control dependent on the function entry (bad() method). \newline

\noindent2. Data Dependence:\newline
- `data' variable is data dependent on `if (privateFive == 5)' statement.\newline
- `KerberosPrincipal principal' is data dependent on the constant string ``test''.\newline
- `KerberosKey key' is data dependent on `principal' and `data' variables.\newline
- `IO.writeLine(key.toString())' is data dependent on `key' variable.
\end{prompt}
We can see that the response w.r.t. program dependency has the same issue as the response w.r.t. data flow described above.

For the total 40 sampled cases, we can observe similar results as the above responses to the code shown in Fig.~\ref{fig:apiseq}.
Therefore, we can conclude that, additional information on data flow is beneficial to enhance the understanding of ChatGPT on the source code.

We also conduct a simple test to investigate whether ChatGPT can extract API call sequences from the function correctly. 
Similar to previous tests, we randomly draw 20 code samples from each dataset and ask ChatGPT the following question:
\begin{prompt}
Please output all the API called by the following code in order. Output format: (xx).(xxx). [CODE]
\end{prompt}

As an example, the response of ChatGPT w.r.t. the code shown in Fig.~\ref{fig:apiseq} is shown below:
\begin{prompt}
\noindent\textbf{The answer for API call sequence}:\newline
1. KerberosPrincipal.constructor\newline
2. KerberosKey.constructor\newline
3. IO.writeLine
\end{prompt}

From the above example, we can observe that ChatGPT can only identify obvious and simple API calls. 
The calls passed as function parameters are not identified by ChatGPT (i.e., String.toCharArray and KerberosKey.toString).
Besides, based on the above response, it is unclear whether ChatGPT can recognize the order of API calls (i.e.,  call KerberosKey.toString first and then call IO.writeLine). 
Therefore, we can conclude that explicitly adding information on API call sequences can benefit ChatGPT and improve its ability to comprehend source code. 
For the total 40 sampled cases, we have a similar observation as the above example.

\subsubsection{Prompts with DFG and API Calls}
\label{sec:dfapi}

Since ChatGPT cannot fully understand \underline{d}ata flow and \underline{A}PI call sequences of functions, based on the role-based prompt $\mathbf{P}_{\text{r-b}}$, we design the following prompts (\textbf{P}$_{\text{r-b-d}}$ and \textbf{P}$_{\text{r-a-b}}$) equipped with auxiliary information on data flow or API calls:
\begin{prompt}{\textbf{P}$_{\text{r-b-d}}$:}
I want you to act as a vulnerability detection system. I will provide you with the original program and the data flow information, and you will act upon them. Is the following program buggy? [CODE]. [DF description].
\end{prompt}
\begin{prompt}{\textbf{P}$_{\text{r-a-b}}$:}
I want you to act as a vulnerability detection system. I will provide you with the original program and the API call sequence, and you will act upon them. [API description]. Is the following program buggy? [CODE].
\end{prompt}
\noindent where [API description] and [DF description] are constructed based on the extracted API call sequences and DFGs described in Sec.~\ref{sec:feature}.
Note that the position of the letter ``d'' and letter ``a'' in the subscripts of \textbf{P}$_{\text{r-b-d}}$ and \textbf{P}$_{\text{r-a-b}}$ indicate the positon of [DF description] and [API description] w.r.t. the part of the task role (``r'') and the part of the basic prompt (``b'').
There are other options for the order but \textbf{P}$_{\text{r-b-d}}$ and \textbf{P}$_{\text{r-a-b}}$ yield better performance than alternatives and we adopt them as the default order for prompts with information on data flow/API calls.
We also discuss the impact of the order in Sec.~\ref{sec:exp_location}.

To obtain [API description], once we obtain the API call sequence of each function (see Sec.~\ref{sec:api}), we can concatenate each function call with specific words.
We propose a template for each API call sequence to construct [API description]: 
\begin{prompt}
The program first calls $a_{1}$, then calls $a_{2}$, $\cdots$, then calls $a_{i}$, $\cdots$, and finally calls $a_{n}$.
\end{prompt}
\noindent where $n$ denotes the number of function calls in the API call sequence.
For example, the [API description] for the API call sequence in Fig.~\ref{fig:apiseq} is:
\begin{prompt}
The program first calls KerberosPrincipal.new, then calls\\String.toCharArray, then calls KerberosKey.new, then calls KerberosKey.toString, and finally calls IO.writeLine.
\end{prompt}

Similarly, we can construct [DF description] for each function based on the extracted tuple described in Sec.~\ref{sec:dfg}, we design a prompt template to integrate DFG into the inputs to ChatGPT:
\begin{prompt}
The data value of the variable $v_i$ at the $p_i$th token comes from/is computed by the variable $v_j$ at the $p_j$th token.
\end{prompt}

For example, the prompt of the data flow in Fig.~\ref{fig:apiseq} is: 
\begin{prompt}
The data value of the variable data at the 17th token comes from the variable data at the 11th token or the variable data at 14th token. The data value of the variable data at the 11th token is computed by the ``7e5tc4s3'' at the 12th token...
\end{prompt}

\subsection{Chain-of-Thought Prompting}
\label{sec:cot}

One remarkable improvement of LLMs over traditional small language models is that LLMs can solve complex tasks via multiple reasoning steps.
The key is the chain-of-thought prompting~\cite{Wei0SBIXCLZ22}.
A chain of thought is a coherent series of intermediate natural language reasoning steps that lead to the final answer for a task.
Since ChatGPT can memorize multi-step interactions, we can attempt chain-of-thought prompting (i.e., design multiple prompts) for vulnerability detection. 

\begin{table*}[t]
\caption{Performance of ChatGPT on Vulnerability Detection using Different Prompts. Best results are shown in bold.}
\label{tab:main}
\resizebox{0.99\textwidth}{!}{
\begin{tabular}{c|ccccccccc|ccccccccc}
\hline
 & \multicolumn{9}{c|}{Java} & \multicolumn{9}{c}{C/C++} \\ \cline{2-19} 
 & \multicolumn{4}{c|}{Vulnerable} & \multicolumn{4}{c|}{Non-Vulnerable} & All & \multicolumn{4}{c|}{Vulnerable} & \multicolumn{4}{c|}{Non-Vulnerable} & All \\ \hline
Prompt & P & R & F1 & \multicolumn{1}{c|}{Det.} & P & R & F1 & \multicolumn{1}{c|}{Det.} & Acc & P & R & F1 & \multicolumn{1}{c|}{Det.} & P & R & F1 & \multicolumn{1}{c|}{Det.} & Acc \\ \hline

CFGNN & \multicolumn{1}{c|}{0.831} & \multicolumn{1}{c|}{0.144} & \multicolumn{1}{c|}{0.245} & \multicolumn{1}{c|}{143/995} & \multicolumn{1}{c|}{0.383} & \multicolumn{1}{c|}{0.948} & \multicolumn{1}{c|}{0.546} & \multicolumn{1}{c|}{529/558} & 0.437 & \multicolumn{1}{c|}{1.000} & \multicolumn{1}{c|}{0.105} & \multicolumn{1}{c|}{0.191} & \multicolumn{1}{c|}{2/19} & \multicolumn{1}{c|}{0.393} & \multicolumn{1}{c|}{1.000} & \multicolumn{1}{c|}{0.564} & \multicolumn{1}{c|}{11/11} & 0.433 \\ \hline

Bugram & \multicolumn{1}{c|}{0.469} & \multicolumn{1}{c|}{0.032} & \multicolumn{1}{c|}{0.061} & \multicolumn{1}{c|}{38/1171} & \multicolumn{1}{c|}{0.435} & \multicolumn{1}{c|}{0.953} & \multicolumn{1}{c|}{0.598} & \multicolumn{1}{c|}{874/917} & 0.436 & \multicolumn{1}{c|}{0.507} & \multicolumn{1}{c|}{0.514} & \multicolumn{1}{c|}{0.51} & \multicolumn{1}{c|}{511/995} & \multicolumn{1}{c|}{0.459} & \multicolumn{1}{c|}{0.453} & \multicolumn{1}{c|}{0.456} & \multicolumn{1}{c|}{411/908} & 0.484 \\ \hline
 
P$_{\text{b}}$ & \multicolumn{1}{c|}{0.659} & \multicolumn{1}{c|}{0.932} & \multicolumn{1}{c|}{0.772} & \multicolumn{1}{c|}{1091/1171} & \multicolumn{1}{c|}{0.815} & \multicolumn{1}{c|}{0.384} & \multicolumn{1}{c|}{0.522} & \multicolumn{1}{c|}{352/917} & 0.691 & \multicolumn{1}{c|}{0.524} & \multicolumn{1}{c|}{0.992} & \multicolumn{1}{c|}{0.686} & \multicolumn{1}{c|}{1007/1015} & \multicolumn{1}{c|}{0.429} & \multicolumn{1}{c|}{0.007} & \multicolumn{1}{c|}{0.013} & \multicolumn{1}{c|}{6/922} & 0.523 \\ \hline
P$_{\text{r-r-b}}$ & \multicolumn{1}{c|}{0.727} & \multicolumn{1}{c|}{0.450} & \multicolumn{1}{c|}{0.556} & \multicolumn{1}{c|}{527/1171} & \multicolumn{1}{c|}{0.528} & \multicolumn{1}{c|}{0.784} & \multicolumn{1}{c|}{0.631} & \multicolumn{1}{c|}{719/917} & 0.597 & \multicolumn{1}{c|}{0.500} & \multicolumn{1}{c|}{0.136} & \multicolumn{1}{c|}{0.214} & \multicolumn{1}{c|}{138/1015} & \multicolumn{1}{c|}{0.472} & \multicolumn{1}{c|}{0.850} & \multicolumn{1}{c|}{0.607} & \multicolumn{1}{c|}{784/922} & 0.476 \\ \hline
P$_{\text{r-b}}$ & \multicolumn{1}{c|}{0.726} & \multicolumn{1}{c|}{0.817} & \multicolumn{1}{c|}{0.769} & \multicolumn{1}{c|}{957/1171} & \multicolumn{1}{c|}{0.722} & \multicolumn{1}{c|}{0.606} & \multicolumn{1}{c|}{0.659} & \multicolumn{1}{c|}{556/917} & 0.725 & \multicolumn{1}{c|}{0.523} & \multicolumn{1}{c|}{0.982} & \multicolumn{1}{c|}{0.682} & \multicolumn{1}{c|}{1009/1015} & \multicolumn{1}{c|}{0.523} & \multicolumn{1}{c|}{0.002} & \multicolumn{1}{c|}{0.004} & \multicolumn{1}{c|}{2/922} & 0.522 \\ \hline
P$_{\text{r-a-b}}$ & \multicolumn{1}{c|}{0.894} & \multicolumn{1}{c|}{0.622} & \multicolumn{1}{c|}{0.734} & \multicolumn{1}{c|}{728/1171} & \multicolumn{1}{c|}{0.652} & \multicolumn{1}{c|}{0.906} & \multicolumn{1}{c|}{0.759} & \multicolumn{1}{c|}{831/917} & \textbf{0.747} & \multicolumn{1}{c|}{0.521} & \multicolumn{1}{c|}{0.758} & \multicolumn{1}{c|}{0.618} & \multicolumn{1}{c|}{646/852} & \multicolumn{1}{c|}{0.490} & \multicolumn{1}{c|}{0.250} & \multicolumn{1}{c|}{0.331} & \multicolumn{1}{c|}{198/791} & 0.514 \\ \hline
P$_{\text{r-b-d}}$ & \multicolumn{1}{c|}{0.784} & \multicolumn{1}{c|}{0.564} & \multicolumn{1}{c|}{0.656} & \multicolumn{1}{c|}{661/1171} & \multicolumn{1}{c|}{0.590} & \multicolumn{1}{c|}{0.802} & \multicolumn{1}{c|}{0.680} & \multicolumn{1}{c|}{735/917} & 0.669 & \multicolumn{1}{c|}{0.525} & \multicolumn{1}{c|}{0.970} & \multicolumn{1}{c|}{0.682} & \multicolumn{1}{c|}{985/1015} & \multicolumn{1}{c|}{0.516} & \multicolumn{1}{c|}{0.035} & \multicolumn{1}{c|}{0.065} & \multicolumn{1}{c|}{32/922} & \textbf{0.525} \\ \hline
P$_{\text{r-a-b-d}}$ & \multicolumn{1}{c|}{0.636} & \multicolumn{1}{c|}{0.735} & \multicolumn{1}{c|}{0.682} & \multicolumn{1}{c|}{861/1171} & \multicolumn{1}{c|}{0.578} & \multicolumn{1}{c|}{0.462} & \multicolumn{1}{c|}{0.514} & \multicolumn{1}{c|}{424/917} & 0.615 & \multicolumn{1}{c|}{0.564} & \multicolumn{1}{c|}{0.159} & \multicolumn{1}{c|}{0.249} & \multicolumn{1}{c|}{136/852} & \multicolumn{1}{c|}{0.489} & \multicolumn{1}{c|}{0.867} & \multicolumn{1}{c|}{0.626} & \multicolumn{1}{c|}{686/791} & 0.500 \\ \hline
\end{tabular}
}
\end{table*}

\vspace{5pt}
\noindent\textbf{Step 1:} 
Intuitively, LLMs can correctly determine whether the code snippet is vulnerable or not, providing that LLMs can first understand the purpose of the code precisely. 
Consequently, we design the first-step prompt for the \textit{intention} of the test code:
\begin{prompt}{\textbf{P}$_{1}^{\text{(chain)}}$:}
Please describe the intent of the given code. [CODE].
\end{prompt}

\vspace{5pt}
\noindent\textbf{Step 2:} 
\noindent Following the first step, we can continue to ask ChatGPT about the vulnerability of the test function.
For instance, we can adopt the role-based basic prompt ($\mathbf{P}_{\text{r-b}}$) in Sec.~\ref{sec:basic prompt} as the second step:

\begin{prompt}{\textbf{P}$_{2,\text{r-b}}^{\text{(chain)}}$:}
I want you to act as a vulnerability detection system. Is the above program buggy? Please answer Yes or No.
\end{prompt}

\vspace{5pt}
\noindent\textbf{Step 2 with Auxiliary Information:} 
Similar to \textbf{P}$_{\text{r-a-b}}$/\textbf{P}$_{\text{r-b-d}}$ proposed in Sec.~\ref{sec:dfapi}, we can also enrich the prompt in the second step with additional information on API calls and data flow:
\begin{prompt}{\textbf{P}$_{2,\,\text{aux}}^{\text{(chain)}}$:}
I want you to act as a vulnerability detection system. Is the above code buggy? Only answer Yes or No. Here is its \uline{API call sequence/data flow information} that you may use: [API description]/[DF description].
\end{prompt}

\section{Experimental Results}
\label{sec:res}

In this section, we report and analyze the experimental results in order to answer the following research questions (RQ):
\begin{itemize}[leftmargin=12pt,topsep=1pt,itemsep=0.3pt]
\item \textbf{RQ1:} Can ChatGPT detect software vulnerability with the help of basic prompts?

\item \textbf{RQ2:} Can API calls and data flow information enhance the prompting?

\item \textbf{RQ3:} Does chain-of-thought prompting affect the accuracy of vulnerability detection?

\item \textbf{RQ4:} Does the order of compositions of the prompt affect the detection?

\item \textbf{RQ5:} How does ChatGPT perform on different vulnerability types?

\end{itemize}

\subsection{Experimental Settings}

\subsubsection{Vulnerability Detection Baselines}
\label{sec:baseline}

In our study, we compared ChatGPT with two state-of-the-art vulnerability detection methods:
\begin{itemize}[leftmargin=12pt,topsep=1pt,itemsep=0.3pt]

\item \textbf{CFGNN}\footnote{\url{https://github.com/zhangj111/ConditionBugs}}~\cite{ZhangWZ0LHL23} is the state-of-the-art condition-based bug detection method, which utilizes API knowledge and CFG-based Graph Neural Network (CFGNN) to detect condition-related bugs automatically.

\item \textbf{Bugram}~\cite{WangCMT16} adopts n-gram language models instead of rules to detect bugs. 
It first models program tokens sequentially using the n-gram language model and then ranks token sequences according to their probabilities in the learned model.
Low-probability sequences are marked as potential bugs.

\end{itemize}
While there are other software vulnerability detection methods, they are designed for one or a few specific vulnerability types.
We choose the above two baselines because they are general detection methods that can cover different types of vulnerability types.
We use the default configurations of baselines in our study.

\subsubsection{Evaluation Metrics}
\label{sec:metric}

We use Accuracy (Acc), Precision (P), Recall (R) and F1 score (F1), which are commonly used to evaluate vulnerability detection methods, to measure the performance of ChatGPT.

\subsection{Effectiveness of Basic Prompts (RQ1)}

Tab.~\ref{tab:main} illustrates the detection performance of ChatGPT using different prompts.
In Tab.~\ref{tab:main}, we report the performance on the vulnerable sample set and the non-vulnerable sample set separately.
We also provide the results on the complete sample set which are denoted by ``All'' in Tab.~\ref{tab:main}.
In Tab.~\ref{tab:main}, ``Det.'' denotes how many test samples are correctly predicted.
For instance, ChatGPT with \textbf{P}$_{\text{b}}$ can correctly predict 1,091 out of 1,171 true vulnerable samples as vulnerable code on the Java dataset.
Note that some test cases are excluded since their features cannot be extracted by baselines/ChatGPT or ChatGPT replies that it cannot handle the task.
Hence, the total number shown in the ``Det.'' column of Tab.~\ref{tab:main} varies in different rows.

\vspace{5pt}
\noindent\textbf{Overall Performance Compared to Baselines.} 
We can see that using basic prompts (\textbf{P}$_{\text{b}}$, \textbf{P}$_{\text{r-b}}$ and \textbf{P}$_{\text{r-r-b}}$), ChatGPT generally outperforms baselines CFGNN and Bugram, 
especially on the Java dataset where the accuracy of ChatGPT with \textbf{P}$_{\text{b}}$ is 58\% and 64\% higher than CFGNN and Bugram, respectively.

\begin{center}
    \begin{tcolorbox}[colback=gray!7, colframe=black, width=8.5cm, arc=1mm, auto outer arc, boxrule=1pt, left=1mm, right=1mm, top = 1mm, bottom = 1mm]{{\textbf{Finding 1:}
    Compared to the two vulnerability detection baselines, ChatGPT shows better performance on vulnerability detection w.r.t. both accuracy and coverage.}}
    \end{tcolorbox}
\end{center}

\vspace{5pt}
\noindent\textbf{Impact of Using a Task Role.}
The documentation provided by OpenAI ~\cite{GBP} for using GPT mentions that the system message can be used to specify the persona used by the model in its replies.
We expect that the prompt design after specifying the task roles will perform better. 
As shown in Tab.~\ref{tab:main}, on the Java dataset, the accuracy of \textbf{P}$_{\text{r-b}}$ is about 5\% higher than the accuracy of \textbf{P}$_{\text{b}}$. 
On the C/C++ dataset, there is a slight increase in the number of correctly detected vulnerabilities and a small decrease in the number of non-vulnerabilities detected, and the overall accuracy remains almost unchanged. 

\begin{center}
    \begin{tcolorbox}[colback=gray!7, colframe=black, width=8.5cm, arc=1mm, auto outer arc, boxrule=1pt, left=1mm, right=1mm, top = 1mm, bottom = 1mm]{{\textbf{Finding 2:}
Including a task role in the prompt may enhance the performance of ChatGPT on vulnerability detection but the improvement is programming-language-specific.
}
}
    \end{tcolorbox}
\end{center}

\vspace{5pt}
\noindent\textbf{Bias of Prediction.} Tab.~\ref{tab:main} also shows that, on both Java and C/C++ data, most functions are predicted to be vulnerable by ChatGPT when using basic prompts. 
Given the unbalanced prediction, it is essential to investigate whether ChatGPT simply predicts most functions as vulnerable without inspecting the details of the source code due to the reason that the question contains the keyword ``buggy''. 
Therefore, we design the reverse-question prompt \textbf{P}$_{\text{r-r-b}}$ and replace ``buggy'' with ``correct'' in the prompt. 
From Tab.~\ref{tab:main}, we can see that using \textbf{P}$_{\text{r-r-b}}$, ChatGPT tends to be biased toward predicting the function as non-vulnerable, suggesting that ChatGPT tends to guess the answer according to the keyword in the prompt.
\begin{center}
    \begin{tcolorbox}[colback=gray!7, colframe=black, width=8.5cm, arc=1mm, auto outer arc, boxrule=1pt, left=1mm, right=1mm, top = 1mm, bottom = 1mm]{{\textbf{Finding 3:}
    With the simple basic prompt, the response of ChatGPT is biased toward the keyword in the prompt.}}
    \end{tcolorbox}
\end{center}

\begin{figure}[t]
    \centering
    \includegraphics[width=1.1\linewidth]{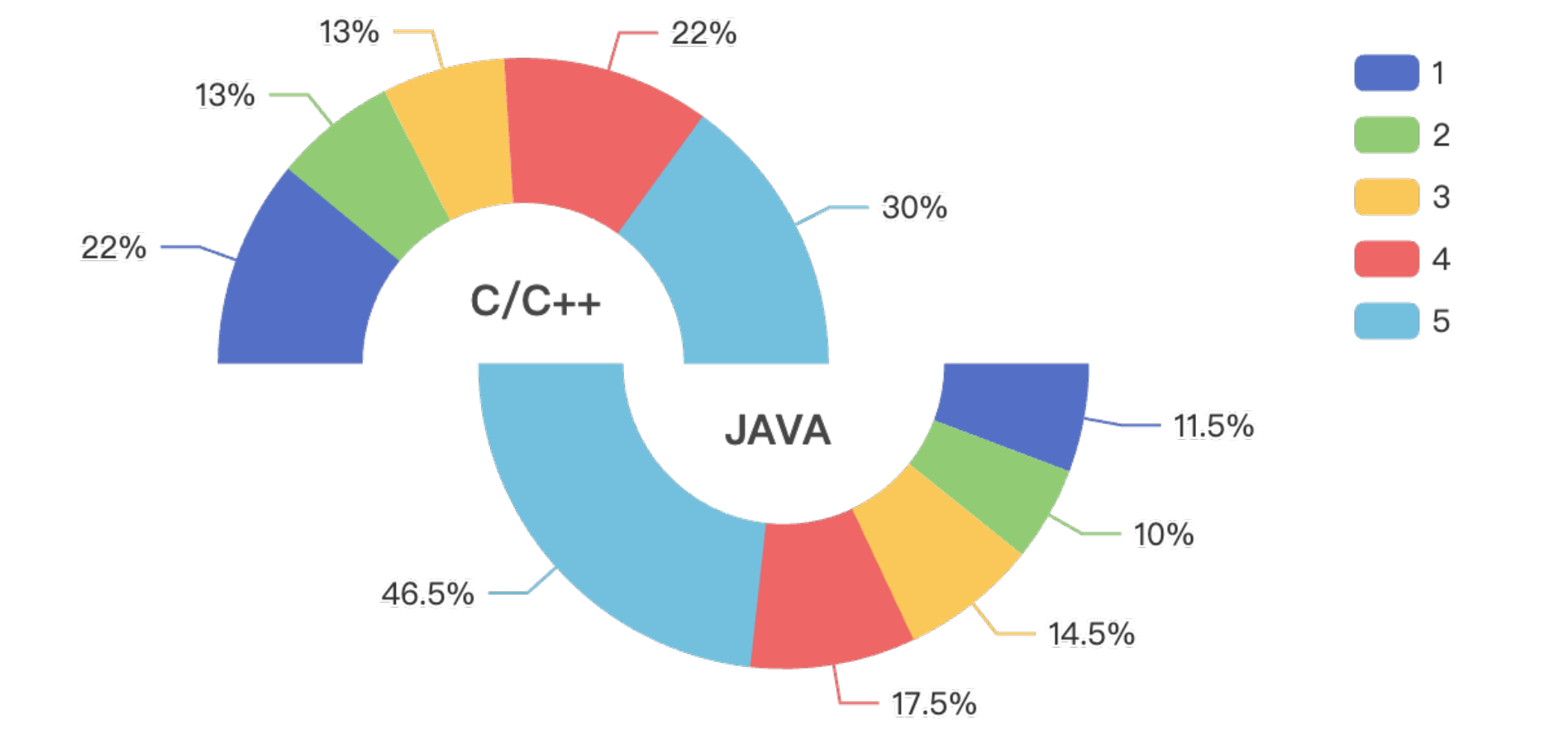}
    \caption{Rating distributions for vulnerability explanations.}
    \label{fig:bug explanation}
\end{figure}

\vspace{5pt}
\noindent\textbf{Comprehension of Vulnerability.}
To investigate whether ChatGPT with basic prompts can correctly understand vulnerabilities. 
We draw 200 samples from the true positive results of vulnerable samples (100 from each dataset) using $\mathbf{P}_{\text{b}}$ and ask ChatGPT the following question following the prompt $\mathbf{P}_{\text{b}}$:
\begin{prompt}
If the answer is yes, please describe the vulnerability.
\end{prompt}
\noindent We invited five master students majoring in Computer Science to manually assess the quality of the 200 vulnerability explanations by giving each explanation a score on a five-point scale. 
The result is reported in Fig.~\ref{fig:bug explanation}.
We can see that high scores (4 or 5) account for 52\% and 64\% on the C/C++ dataset and the Java dataset, respectively. 
In other words, ChatGPT with the basic prompt \textbf{P}$_{\text{b}}$ does not understand almost half of the vulnerabilities it ``detects''.
The observation can be the reason for the detection bias in Finding 3.

\begin{center}
    \begin{tcolorbox}[colback=gray!7, colframe=black, width=8.5cm, arc=1mm, auto outer arc, boxrule=1pt, left=1mm, right=1mm, top = 1mm, bottom = 1mm]{{\textbf{Finding 4:}
With the basic prompt, ChatGPT is more capable of identifying vulnerabilities in Java programs than in C/C++ programs, but it cannot understand vulnerabilities comprehensively.
}

}
    \end{tcolorbox}
\end{center}

\begin{table*}[t]
\caption{Results of chain-of-thought prompting. Best results are shown in bold.}
\label{tab:cot}
\begin{center}
\resizebox{0.99\textwidth}{!}{
\begin{tabular}{c|ccccccccc|ccccccccc}
\hline
              & \multicolumn{9}{c|}{Java}                                                                                                                                                                                                                   & \multicolumn{9}{c}{C/C++}                                                                                                                                                                                                                          \\ \cline{2-19} 
              & \multicolumn{4}{c|}{Vulnerable}                                                                                        & \multicolumn{4}{c|}{Non-Vulnerable}                                                                                    & All   & \multicolumn{4}{c|}{Vulnerable}                                                                                           & \multicolumn{4}{c|}{Non-Vulnerable}                                                                                       & All    \\ \hline
Prompt        & P                          & R                          & F1                         & \multicolumn{1}{c|}{Det.} & P                          & R                          & F1                         & \multicolumn{1}{c|}{Det.} & Acc   & P                           & R                           & F1                          & \multicolumn{1}{c|}{Det.} & P                           & R                           & F1                          & \multicolumn{1}{c|}{Det.} & Acc    \\ \hline
P$_{\text{2, r-b}}^{\text{(chain)}}$    & \multicolumn{1}{c|}{0.668} & \multicolumn{1}{c|}{0.93}  & \multicolumn{1}{c|}{0.777} & \multicolumn{1}{c|}{1089/1171} & \multicolumn{1}{c|}{0.821} & \multicolumn{1}{c|}{0.409} & \multicolumn{1}{c|}{0.546} & \multicolumn{1}{c|}{375/917}  & \textbf{0.701}& \multicolumn{1}{c|}{0.675} & \multicolumn{1}{c|}{0.973} & \multicolumn{1}{c|}{0.797} & \multicolumn{1}{c|}{988/1015}  & \multicolumn{1}{c|}{0.943} & \multicolumn{1}{c|}{0.485} & \multicolumn{1}{c|}{0.640} & \multicolumn{1}{c|}{447/922}  & \textbf{0.741} \\ \hline
P$_{\text{2, r-a-b}}^{\text{(chain)}}$   & \multicolumn{1}{c|}{0.688} & \multicolumn{1}{c|}{0.703} & \multicolumn{1}{c|}{0.695} & \multicolumn{1}{c|}{823/1171}  & \multicolumn{1}{c|}{0.609} & \multicolumn{1}{c|}{0.592} & \multicolumn{1}{c|}{0.601} & \multicolumn{1}{c|}{543/917}  & 0.654 & \multicolumn{1}{c|}{0.661} & \multicolumn{1}{c|}{0.855} & \multicolumn{1}{c|}{0.746} & \multicolumn{1}{c|}{728/852}  & \multicolumn{1}{c|}{0.771} & \multicolumn{1}{c|}{0.527} & \multicolumn{1}{c|}{0.626} & \multicolumn{1}{c|}{417/791}  & 0.697 \\ \hline
P$_{\text{2, r-b-d}}^{\text{(chain)}}$   & \multicolumn{1}{c|}{0.704} & \multicolumn{1}{c|}{0.651} & \multicolumn{1}{c|}{0.676} & \multicolumn{1}{c|}{762/1171}  & \multicolumn{1}{c|}{0.593} & \multicolumn{1}{c|}{0.65}  & \multicolumn{1}{c|}{0.620}  & \multicolumn{1}{c|}{596/917}  & 0.65  & \multicolumn{1}{c|}{0.794} & \multicolumn{1}{c|}{0.614} & \multicolumn{1}{c|}{0.692} & \multicolumn{1}{c|}{623/1015}  & \multicolumn{1}{c|}{0.660} & \multicolumn{1}{c|}{0.824} & \multicolumn{1}{c|}{0.733} & \multicolumn{1}{c|}{760/922}  & 0.714 \\ \hline
P$_{\text{2, r-a-b-d}}^{\text{(chain)}}$ & \multicolumn{1}{c|}{0.714} & \multicolumn{1}{c|}{0.545} & \multicolumn{1}{c|}{0.618} & \multicolumn{1}{c|}{638/1171}  & \multicolumn{1}{c|}{0.554} & \multicolumn{1}{c|}{0.721} & \multicolumn{1}{c|}{0.626} & \multicolumn{1}{c|}{661/917}  & 0.622 & \multicolumn{1}{c|}{0.814} & \multicolumn{1}{c|}{0.589} & \multicolumn{1}{c|}{0.684} & \multicolumn{1}{c|}{502/852}  & \multicolumn{1}{c|}{0.659} & \multicolumn{1}{c|}{0.855} & \multicolumn{1}{c|}{0.744} & \multicolumn{1}{c|}{676/791}  & 0.717 \\ \hline
\end{tabular}
}
\end{center}
\end{table*}
\subsection{Usefulness of Auxiliary Information (RQ2)}
Next, we investigate the impact of incorporating auxiliary information in the prompt.
It can be seen from Tab.~\ref{tab:main} that, on both datasets, the accuracy of \textbf{P}$_{\text{r-a-b}}$ and \textbf{P}$_{\text{r-d-b}}$ decreases on detecting vulnerable data compared to \textbf{P}$_{\text{r-b}}$.
On the non-vulnerability data, the accuracy of \textbf{P}$_{\text{r-a-b}}$ and \textbf{P}$_{\text{r-d-b}}$ significantly increases. 
We can also see that, on the Java dataset, adding API calls to the prompt results in an accuracy of 0.747, which is about 8\% higher than \textbf{P}$_{\text{r-b}}$. 
On the C/C++ dataset, the prompt including data flow information works best but it does not exceed \textbf{P}$_{\text{b}}$ by a large margin.

\begin{center}
    \begin{tcolorbox}[colback=gray!7, colframe=black, width=8.5cm, arc=1mm, auto outer arc, boxrule=1pt, left=1mm, right=1mm, top = 1mm, bottom = 1mm]{{\textbf{Finding 5:} 
    Different auxiliary information has diverse effects on different programming languages. Incorporating API calls is more effective for detecting vulnerability in Java functions, while including data flow information contributes a little to the understanding of C/C++ vulnerable programs.
    }}
    \end{tcolorbox}
\end{center}
\subsection{Effectiveness of Chain-of-Thought Prompting (RQ3)}
\subsubsection{Performance of Vulnerability Detection in Step 2.}
In Tab.~\ref{tab:cot}, we report the performance of vulnerability detection of step 2 in chain-of-thought prompting. 
We can see that \textbf{P}$_{2,\text{r-b}}^{(\text{chain})}$ achieves the best accuracy on both Java and C/C++ dataset. 
Surprisingly, \textbf{P}$_{2,\text{r-a-b}}^{(\text{chain})}$ and \textbf{P}$_{2,\text{r-b-d}}^{(\text{chain})}$, which incorporate auxiliary information, perform worse than \textbf{P}$_{2,\text{r-b}}^{(\text{chain})}$ regarding accuracy.
Taking a closer look, we find that adding information on API calls or data flow causes ChatGPT to identify more non-vulnerable samples correctly. 
The number of detected non-vulnerable samples is increasing while the detected vulnerable ones are decreasing as auxiliary information is added. 
Besides, for the Java dataset, the results of chain-of-thought prompting are worse than the basic prompts shown in Tab.~\ref{tab:main}. 
However, for C/C++ dataset, all the chain-of-thought prompts achieve superior results than basic prompts. 
We speculate that the difference is caused by the position of extra information and code in the prompt and we will investigate the impact of the position of the compositions of prompts in Sec.~\ref{sec:exp_location}.
\begin{center}
    \begin{tcolorbox}[colback=gray!7, colframe=black, width=8.5cm, arc=1mm, auto outer arc, boxrule=1pt, left=1mm, right=1mm, top = 1mm, bottom = 1mm]{{\textbf{Finding 6:} Chain-of-thought prompting has different effects on Java and C/C++. It provides a great improvement on C/C++ dataset but degrades the detection performance on Java dataset.}}
    \end{tcolorbox}
\end{center}

\subsubsection{Impacts of Including the Summary of Code.}
\begin{figure}[t]
    \centering
    \includegraphics[width=1\linewidth]{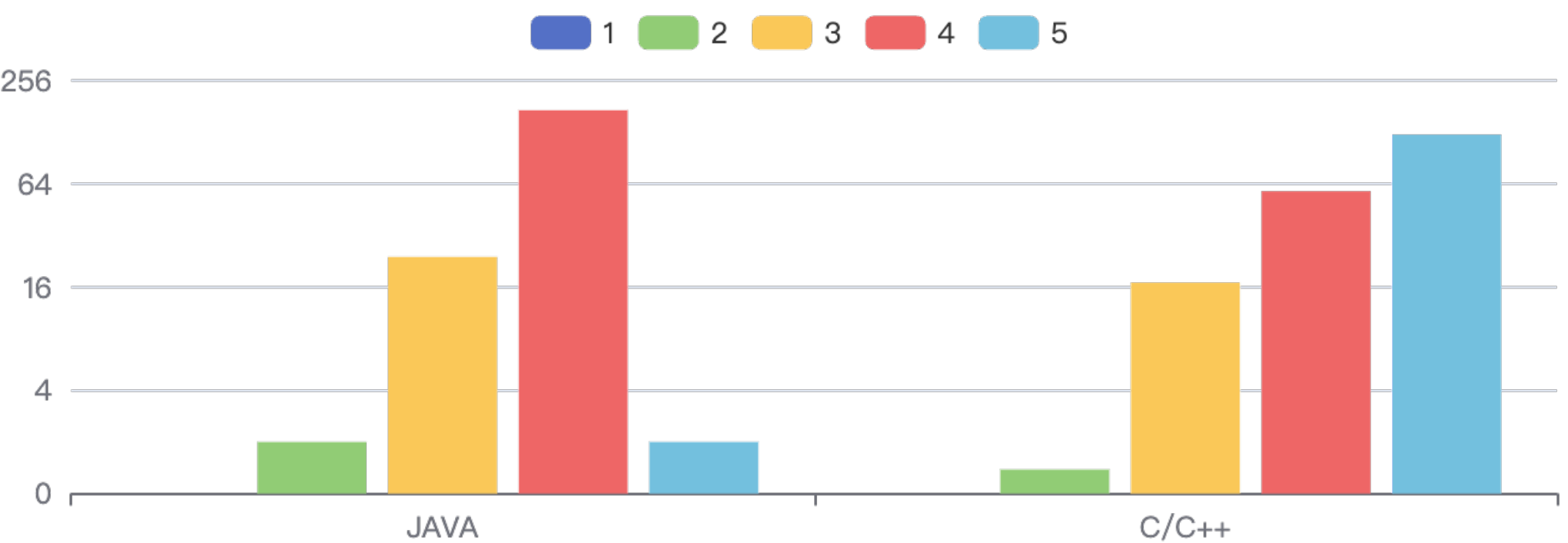}
    \caption{Rating distributions for sampled answers to step 1.}
    \label{fig:sam}
\end{figure}

The intermediate step in chain-of-thought prompting affects the quality of the final answer.
Although we focus on the detection performance of step 2 (i.e., \textbf{P}$_{2}^{\text{(chain)}}$) in the chain-of-thought prompting, we also take a step to check the answer in step 1. 
We randomly selected 200 answers to \textbf{P}$_{1}^{\text{(chain)}}$ generated by ChatGPT.
The answers correspond to 100 vulnerable test code and 100 non-vulnerable test code randomly sampled from both Java and C/C++ data. 
We ask five master students majoring in Computer Science to give a rating using 5 point rating scale to each answer to assess whether ChatGPT can understand the functionality of the code.
The distribution of ratings is shown in Fig.~\ref{fig:sam}.
We can see that there are no answers that are given a rating of 1 and nearly 90\% of the answers have a score of 4 or higher. 
The result demonstrates that ChatGPT can understand the purpose of the function precisely. 

\begin{center}
    \begin{tcolorbox}[colback=gray!7, colframe=black, width=8.5cm, arc=1mm, auto outer arc, boxrule=1pt, left=1mm, right=1mm, top = 1mm, bottom = 1mm]{{\textbf{Finding 7:} ChatGPT can accurately understand the functionality of code in vulnerability detection}.}
    \end{tcolorbox}
\end{center}

Chain-of-thought prompting demands the model
to summarize the code first and then detect the vulnerability
Alternatively, we can regard the code summary as auxiliary information like API calls and data flow to construct a single prompt instead of two-step chain-of-thought prompts.
Due to the restriction of input length of ChatGPT, we only test basic prompts that do not contain API calls and data flow. 
We use \textit{$s$} to represent the code summary in the prompt.
The results using summary-aided prompts are shown in Tab.~\ref{tab:summary}.
We can see that the accuracy decreases after adding code summary to the prompt to detect vulnerability on the Java dataset.
Differently, the performance increases on C/C++ dataset.
\begin{center}
    \begin{tcolorbox}[colback=gray!7, colframe=black, width=8.5cm, arc=1mm, auto outer arc, boxrule=1pt, left=1mm, right=1mm, top = 1mm, bottom = 1mm]{{\textbf{Finding 8:} Adding high-quality code summaries to prompts can improve the detection performance of ChatGPT. But the impact is programming-language-specific.}}
    \end{tcolorbox}
\end{center}

\subsection{Impacts of Position (RQ4)}
\label{sec:exp_location}

\begin{table}[t]
\caption{Accuracy before/after adding code summary.}
\label{tab:summary}
\begin{tabular}{c|cc|cc}
\hline
 & \textbf{P}$_{\text{b}}$ & \textbf{P}$_{\text{r-b}}$ & \textbf{P}$_{\text{s-b}}$ & \textbf{P}$_{\text{r-s-b}}$ \\ \hline
Java & 0.691 & 0.725 & 0.657 & 0.671 \\
C/C++ & 0.523 & 0.522 & 0.526 & 0.531 \\ \hline
\end{tabular}
\end{table}

\begin{figure*}[t]
    \centering
    \includegraphics[width=1\linewidth]{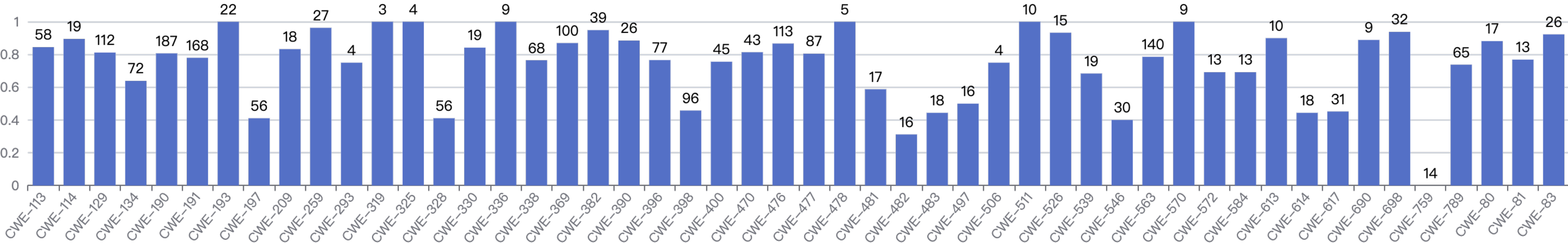}
    \caption{Accuracy of different vulnerability types on the Java dataset. y coordinate indicates the accuracy and the numbers above the bars denote the numbers of samples in each vulnerability type.}
    \label{fig:bar_Java}
    \vspace{10pt}
\end{figure*}

\begin{table*}[t]
\caption{Results for using different settings of position in the prompts. Best results are shown in bold.}
\label{tab:order}
\begin{center}
\resizebox{0.99\textwidth}{!}{
\begin{tabular}{c|ccccccccc|ccccccccc}
\hline
            & \multicolumn{9}{c|}{Java}                                                                                     & \multicolumn{9}{c}{C/C++}                                                                                            \\ \cline{2-19} 
            & \multicolumn{4}{c|}{Vulnerable}                         & \multicolumn{4}{c|}{Non-Vulnerable}                     & All   & \multicolumn{4}{c|}{Vulnerable}                            & \multicolumn{4}{c|}{Non-Vulnerable}                        & All    \\ \hline
Prompt      & P     & R     & F1    & \multicolumn{1}{c|}{Det.} & P     & R     & F1    & \multicolumn{1}{c|}{Det.} & Acc   & P      & R      & F1     & \multicolumn{1}{c|}{Det.} & P      & R      & F1     & \multicolumn{1}{c|}{Det.} & Acc    \\ \hline
P$_{\text{b-a}}$   & 0.623 & 0.832 & 0.712 & \multicolumn{1}{c|}{974/1171}  & 0.624 & 0.357 & 0.454 & \multicolumn{1}{c|}{327/917}  & 0.623 & 0.480 & 0.028 & 0.053 & \multicolumn{1}{c|}{24/852}   & 0.480 & 0.967 & 0.642 & \multicolumn{1}{c|}{765/791}  & 0.480 \\
P$_{\text{a-b}}$   & 0.832 & 0.435 & 0.571 & \multicolumn{1}{c|}{509/1171}  & 0.551 & 0.888 & 0.680  & \multicolumn{1}{c|}{814/917}  & 0.634 & 0.518 & 0.532 & 0.525 & \multicolumn{1}{c|}{453/852}  & 0.481 & 0.467 & 0.473 & \multicolumn{1}{c|}{369/791}  & 0.500 \\ \hline
P$_{\text{r-b-a}}$ & 0.721 & 0.687 & 0.704 & \multicolumn{1}{c|}{805/1171}  & 0.623 & 0.660  & 0.641 & \multicolumn{1}{c|}{605/917}  & 0.675 & 0.519 & 0.965 & 0.675 & \multicolumn{1}{c|}{822/852}  & 0.492 & 0.037 & 0.068 & \multicolumn{1}{c|}{29/791}   & 0.518 \\
P$_{\text{r-a-b}}$ & 0.894 & 0.622 & 0.734 & \multicolumn{1}{c|}{728/1171}  & 0.652 & 0.906 & 0.759 & \multicolumn{1}{c|}{831/917}  & \textbf{0.747} & 0.521 & 0.758 & 0.618 & \multicolumn{1}{c|}{646/852}  & 0.490 & 0.250 & 0.331 & \multicolumn{1}{c|}{198/791}  & 0.514 \\ \hline
P$_{\text{b-d}}$   & 0.683 & 0.754 & 0.717 & \multicolumn{1}{c|}{883/1171}  & 0.638 & 0.554 & 0.593 & \multicolumn{1}{c|}{508/917}  & 0.666 & 0.522 & 0.943 & 0.672 & \multicolumn{1}{c|}{957/1015}  & 0.442 & 0.050 & 0.090 & \multicolumn{1}{c|}{46/922}   & 0.518 \\
P$_{\text{d-b}}$   & 0.648 & 0.438 & 0.523 & \multicolumn{1}{c|}{513/1171}  & 0.492 & 0.696 & 0.577 & \multicolumn{1}{c|}{638/917}  & 0.551 & 0.536 & 0.610 & 0.571 & \multicolumn{1}{c|}{619/1015}  & 0.494 & 0.420 & 0.415 & \multicolumn{1}{c|}{387/922}  & 0.519 \\ \hline
P$_{\text{r-b-d}}$ & 0.784 & 0.564 & 0.656 & \multicolumn{1}{c|}{661/1171}  & 0.590  & 0.802 & 0.680  & \multicolumn{1}{c|}{735/917}  & 0.669 & 0.525 & 0.970 & 0.682 & \multicolumn{1}{c|}{985/1015}  & 0.516 & 0.035 & 0.065 & \multicolumn{1}{c|}{32/922}   & \textbf{0.525} \\
P$_{\text{r-d-b}}$ & 0.614 & 0.743 & 0.672 & \multicolumn{1}{c|}{870/1171}  & 0.551 & 0.402 & 0.465 & \multicolumn{1}{c|}{369/917}  & 0.593 & 0.502 & 0.742 & 0.599 & \multicolumn{1}{c|}{753/1015}  & 0.401 & 0.190 & 0.258 & \multicolumn{1}{c|}{175/922}  & 0.479 \\ \hline
\end{tabular}
}
\end{center}
\vspace{10pt}
\end{table*}
 
Since there are various options for the positions of different auxiliary information in the prompt, we also investigate how the position of additional information affects the detection. 
We attempt two orders by placing [API description] or [DF description] before or after the source code [CODE] for both basic prompt \textbf{P}$_\text{b}$ and role-based basic prompt \textbf{P}$_\text{r-b}$. 
An example is shown below:
\begin{center}
\fcolorbox{gray!5}{gray!7}{\parbox{\linewidth}
{\textbf{Basic Prompt}: [POS1]. \textbf{P}$_{\text{b}}$. [POS2].\newline
\textbf{Role Prompt}: \textbf{P}$_{\text{r}}$. [POS1]. \textbf{P}$_{\text{b}}$. [POS2].
}}
\end{center}
where [POS1] and [POS2] represent the two positions to place [API description] and [DF description]. 
Tab.~\ref{tab:order} provides the results using different positions. 

We can see that the role-based prompt with [API description] located in [POS1] achieves the best accuracy. 
Specifically, we can find that the overall accuracy of placing [API description] in [POS1] is higher than that in [POS2] for both basic prompt and role-based prompt. 
But the improvement for the basic prompt is not as significant as for the role-based prompt, suggesting that different prompts have varying sensitivity for the position.

In addition, we can observe that the performance on vulnerable and non-vulnerable data is different when the position of API information varies. 
The number of true positive samples decreases on vulnerable data but increases on non-vulnerable data when placing [API description] in [POS1]. 
The injection of API information seems more conducive to judgment on non-vulnerable data. 
However, the situation is completely different when adding data flow information. 
The overall accuracy of the prompts with data flow information in [POS2] outperforms those placing data flow information in [POS1]. 
Consequently, the position of API calls or data flow information has a great influence on the performance of ChatGPT, and the suitable choice of position varies for different auxiliary information.

\begin{center}
    \begin{tcolorbox}[colback=gray!7, colframe=black, width=8.5cm, arc=1mm, auto outer arc, boxrule=1pt, left=1mm, right=1mm, top = 1mm, bottom = 1mm]{{\textbf{Finding 9:} Prompting can achieve better performance when placing API calls before the code and placing data flow information after the code. In addition, adding API call information contributes more to the correct prediction of non-vulnerable samples, while including data flow information contributes more to the accurate prediction of vulnerable samples.}}
    \end{tcolorbox}
\end{center}

\subsection{Performance on Different Vulnerability Types (RQ5)}
\label{sec:types}

Due to the page limit, we only report the results on the Java dataset when illustrating the detection performance of different vulnerability types.
Similar trends can be observed on the C/C++ dataset.

We show the accuracy for the prompt \textbf{P}$_\text{r-a-b}$ w.r.t. each vulnerability type in Fig.~\ref{fig:bar_Java}. 
We can see that ChatGPT achieves 100\% accuracy on 7 vulnerability types and we find that some types among the 7 perfectly-detect types, such as CWE-193 (Off-by-one Error), CWE-511 (Logic/Time Bomb) and CWE-570 (Expression is Always False), are easy to identify because they have obvious boundary issues or logical errors. 
Besides, ChatGPT can achieve an accuracy of over 0.5 in more than 80\% of vulnerability types (41/50), indicating that ChatGPT can detect many vulnerability types even though some vulnerable cases are wrongly classified.

We also check the types that get a low accuracy score lower than 0.5. For example, none of the samples in CWE-759 (Use of a One-Way Hash without a Salt) are predicted correctly. 
This kind of vulnerability usually occurs when the operation of adding salt is missing. 
We infer that it requires human intervention to set rules or inject extra knowledge and thus it is difficult for ChatGPT to detect solely based on code and the auxiliary information used in this paper. 
CWE-328 (Use of Weak Hash) has a similar issue, but ChatGPT achieves a  better accuracy of 0.411 on CWE-328 than CWE-759. 
Apart from that, ChatGPT performs poorly on CWE-482 (Comparing instead of Assigning) and CWE-546 (Suspicious Comment) as well. 
For CWE-482, the vulnerability occurs when the ``=='' symbol is mistakenly used as ``=''. For example, ``\textit{if((isZero == (zeroOrOne == 0)) == true)}'', which is a line extracted from a sample function in CWE-482, should be ``\textit{(isZero = (zeroOrOne == 0))}''. 
There is no grammatical error in the sample, indicating that ChatGPT cannot distinguish ``=='' and ``='' via understanding the context. 
For CWE-546, it is triggered for suspicious comments, which does not actually affect the functionality of the program. 
Therefore, without processing the comments, ChatGPT cannot predict CWE-546 correctly.

\begin{center}
    \begin{tcolorbox}[colback=gray!7, colframe=black, width=8.5cm, arc=1mm, auto outer arc, boxrule=1pt, left=1mm, right=1mm, top = 1mm, bottom = 1mm]{{\textbf{Finding 10:} ChatGPT can detect vulnerabilities well on grammar-related or some boundary-related vulnerability types. But it can not achieve a high accuracy on types that are basically irrelevant to the context or those that require the model to fully understand the context.}}
    \end{tcolorbox}
\end{center}

\section{Threats of Validity}
\label{sec:threats}

\vspace{5pt}
\noindent\textbf{Version of ChatGPT:} Since ChatGPT is still being actively updated, the experimental results in this paper only represent the performance of ChatGPT 4. 
For future versions, the conclusions in this paper may become invalid.

\vspace{5pt}
\noindent\textbf{Different Programming Languages:} We conduct experiments on both Java and C/C++ datasets. But as observed in this paper, ChatGPT sometimes performs differently on the two programming languages. It is possible that ChatGPT may show different detection performance on vulnerability detection for other programming languages.

\vspace{5pt}
\noindent\textbf{Human Annotator}. We ask some master students majoring in Computer Science to evaluate the quality of vulnerability explanations and code summaries. Due to the difference of knowledge background, the human evaluation may not be completely correct and unbiased.

\vspace{5pt}
\noindent\textbf{Vulnerability Types.}
The number of samples in different vulnerability types varies as shown in Fig.~\ref{fig:bar_Java}. 
Therefore, the findings in this study may not be universal across all vulnerability types.
It is necessary to collect more vulnerability samples to enrich the two datasets to verify the findings on more data.

\section{Conclusion}
\label{sec:con}
LLMs with stunning intelligence have significantly affected various areas.
In this paper, we study the ability of prompt-enhanced ChatGPT in software vulnerability detection, an important task to ensure the security of software. In particular, we design several prompts that are tailored for vulnerability detection, including adding extra information (i.e., API calls and data flow information) and leveraging chain-of-thought prompting. We conduct extensive experiments on two vulnerability datasets to demonstrate the effectiveness of prompt-enhanced vulnerability detection using ChatGPT. Our study also points out the merit and demerit of using ChatGPT for vulnerability detection.

\bibliographystyle{ACM-Reference-Format}
\bibliography{main}


\begin{thebibliography}{79}


\ifx \showCODEN    \undefined \def \showCODEN     #1{\unskip}     \fi
\ifx \showDOI      \undefined \def \showDOI       #1{#1}\fi
\ifx \showISBNx    \undefined \def \showISBNx     #1{\unskip}     \fi
\ifx \showISBNxiii \undefined \def \showISBNxiii  #1{\unskip}     \fi
\ifx \showISSN     \undefined \def \showISSN      #1{\unskip}     \fi
\ifx \showLCCN     \undefined \def \showLCCN      #1{\unskip}     \fi
\ifx \shownote     \undefined \def \shownote      #1{#1}          \fi
\ifx \showarticletitle \undefined \def \showarticletitle #1{#1}   \fi
\ifx \showURL      \undefined \def \showURL       {\relax}        \fi
\providecommand\bibfield[2]{#2}
\providecommand\bibinfo[2]{#2}
\providecommand\natexlab[1]{#1}
\providecommand\showeprint[2][]{arXiv:#2}

\bibitem[Anil et~al\mbox{.}(2023)]%
        {abs-2305-10403}
\bibfield{author}{\bibinfo{person}{Rohan Anil}, \bibinfo{person}{Andrew~M.
  Dai}, \bibinfo{person}{Orhan Firat}, \bibinfo{person}{Melvin Johnson},
  \bibinfo{person}{Dmitry Lepikhin}, \bibinfo{person}{Alexandre Passos},
  \bibinfo{person}{Siamak Shakeri}, \bibinfo{person}{Emanuel Taropa},
  \bibinfo{person}{Paige Bailey}, \bibinfo{person}{Zhifeng Chen},
  \bibinfo{person}{Eric Chu}, \bibinfo{person}{Jonathan~H. Clark},
  \bibinfo{person}{Laurent~El Shafey}, \bibinfo{person}{Yanping Huang},
  \bibinfo{person}{Kathy Meier{-}Hellstern}, \bibinfo{person}{Gaurav Mishra},
  \bibinfo{person}{Erica Moreira}, \bibinfo{person}{Mark Omernick},
  \bibinfo{person}{Kevin Robinson}, \bibinfo{person}{Sebastian Ruder},
  \bibinfo{person}{Yi Tay}, \bibinfo{person}{Kefan Xiao},
  \bibinfo{person}{Yuanzhong Xu}, \bibinfo{person}{Yujing Zhang},
  \bibinfo{person}{Gustavo~Hern{\'{a}}ndez {\'{A}}brego},
  \bibinfo{person}{Junwhan Ahn}, \bibinfo{person}{Jacob Austin},
  \bibinfo{person}{Paul Barham}, \bibinfo{person}{Jan~A. Botha},
  \bibinfo{person}{James Bradbury}, \bibinfo{person}{Siddhartha Brahma},
  \bibinfo{person}{Kevin Brooks}, \bibinfo{person}{Michele Catasta},
  \bibinfo{person}{Yong Cheng}, \bibinfo{person}{Colin Cherry},
  \bibinfo{person}{Christopher~A. Choquette{-}Choo}, \bibinfo{person}{Aakanksha
  Chowdhery}, \bibinfo{person}{Cl{\'{e}}ment Crepy}, \bibinfo{person}{Shachi
  Dave}, \bibinfo{person}{Mostafa Dehghani}, \bibinfo{person}{Sunipa Dev},
  \bibinfo{person}{Jacob Devlin}, \bibinfo{person}{Mark D{\'{\i}}az},
  \bibinfo{person}{Nan Du}, \bibinfo{person}{Ethan Dyer},
  \bibinfo{person}{Vladimir Feinberg}, \bibinfo{person}{Fangxiaoyu Feng},
  \bibinfo{person}{Vlad Fienber}, \bibinfo{person}{Markus Freitag},
  \bibinfo{person}{Xavier Garcia}, \bibinfo{person}{Sebastian Gehrmann},
  \bibinfo{person}{Lucas Gonzalez}, {and} \bibinfo{person}{et al.}}
  \bibinfo{year}{2023}\natexlab{}.
\newblock \showarticletitle{PaLM 2 Technical Report}.
\newblock \bibinfo{journal}{\emph{arXiv Preprint}} (\bibinfo{year}{2023}).
\newblock
\urldef\tempurl%
\url{https://arxiv.org/abs/2305.10403}
\showURL{%
\tempurl}


\bibitem[Black et~al\mbox{.}(2022)]%
        {abs-2204-06745}
\bibfield{author}{\bibinfo{person}{Sid Black}, \bibinfo{person}{Stella
  Biderman}, \bibinfo{person}{Eric Hallahan}, \bibinfo{person}{Quentin
  Anthony}, \bibinfo{person}{Leo Gao}, \bibinfo{person}{Laurence Golding},
  \bibinfo{person}{Horace He}, \bibinfo{person}{Connor Leahy},
  \bibinfo{person}{Kyle McDonell}, \bibinfo{person}{Jason Phang},
  \bibinfo{person}{Michael Pieler}, \bibinfo{person}{USVSN~Sai Prashanth},
  \bibinfo{person}{Shivanshu Purohit}, \bibinfo{person}{Laria Reynolds},
  \bibinfo{person}{Jonathan Tow}, \bibinfo{person}{Ben Wang}, {and}
  \bibinfo{person}{Samuel Weinbach}.} \bibinfo{year}{2022}\natexlab{}.
\newblock \showarticletitle{GPT-NeoX-20B: An Open-Source Autoregressive
  Language Model}.
\newblock \bibinfo{journal}{\emph{arXiv Preprint}} (\bibinfo{year}{2022}).
\newblock
\urldef\tempurl%
\url{https://arxiv.org/abs/2204.06745}
\showURL{%
\tempurl}


\bibitem[Bommasani et~al\mbox{.}(2021)]%
        {abs-2108-07258}
\bibfield{author}{\bibinfo{person}{Rishi Bommasani}, \bibinfo{person}{Drew~A.
  Hudson}, \bibinfo{person}{Ehsan Adeli}, \bibinfo{person}{Russ~B. Altman},
  \bibinfo{person}{Simran Arora}, \bibinfo{person}{Sydney von Arx},
  \bibinfo{person}{Michael~S. Bernstein}, \bibinfo{person}{Jeannette Bohg},
  \bibinfo{person}{Antoine Bosselut}, \bibinfo{person}{Emma Brunskill},
  \bibinfo{person}{Erik Brynjolfsson}, \bibinfo{person}{Shyamal Buch},
  \bibinfo{person}{Dallas Card}, \bibinfo{person}{Rodrigo Castellon},
  \bibinfo{person}{Niladri~S. Chatterji}, \bibinfo{person}{Annie~S. Chen},
  \bibinfo{person}{Kathleen Creel}, \bibinfo{person}{Jared~Quincy Davis},
  \bibinfo{person}{Dorottya Demszky}, \bibinfo{person}{Chris Donahue},
  \bibinfo{person}{Moussa Doumbouya}, \bibinfo{person}{Esin Durmus},
  \bibinfo{person}{Stefano Ermon}, \bibinfo{person}{John Etchemendy},
  \bibinfo{person}{Kawin Ethayarajh}, \bibinfo{person}{Li Fei{-}Fei},
  \bibinfo{person}{Chelsea Finn}, \bibinfo{person}{Trevor Gale},
  \bibinfo{person}{Lauren Gillespie}, \bibinfo{person}{Karan Goel},
  \bibinfo{person}{Noah~D. Goodman}, \bibinfo{person}{Shelby Grossman},
  \bibinfo{person}{Neel Guha}, \bibinfo{person}{Tatsunori Hashimoto},
  \bibinfo{person}{Peter Henderson}, \bibinfo{person}{John Hewitt},
  \bibinfo{person}{Daniel~E. Ho}, \bibinfo{person}{Jenny Hong},
  \bibinfo{person}{Kyle Hsu}, \bibinfo{person}{Jing Huang},
  \bibinfo{person}{Thomas Icard}, \bibinfo{person}{Saahil Jain},
  \bibinfo{person}{Dan Jurafsky}, \bibinfo{person}{Pratyusha Kalluri},
  \bibinfo{person}{Siddharth Karamcheti}, \bibinfo{person}{Geoff Keeling},
  \bibinfo{person}{Fereshte Khani}, \bibinfo{person}{Omar Khattab},
  \bibinfo{person}{Pang~Wei Koh}, \bibinfo{person}{Mark~S. Krass},
  \bibinfo{person}{Ranjay Krishna}, \bibinfo{person}{Rohith Kuditipudi}, {and}
  \bibinfo{person}{et al.}} \bibinfo{year}{2021}\natexlab{}.
\newblock \showarticletitle{On the Opportunities and Risks of Foundation
  Models}.
\newblock \bibinfo{journal}{\emph{arXiv Preprint}} (\bibinfo{year}{2021}).
\newblock
\urldef\tempurl%
\url{https://arxiv.org/abs/2108.07258}
\showURL{%
\tempurl}


\bibitem[Brown et~al\mbox{.}(2020)]%
        {BrownMRSKDNSSAA20}
\bibfield{author}{\bibinfo{person}{Tom~B. Brown}, \bibinfo{person}{Benjamin
  Mann}, \bibinfo{person}{Nick Ryder}, \bibinfo{person}{Melanie Subbiah},
  \bibinfo{person}{Jared Kaplan}, \bibinfo{person}{Prafulla Dhariwal},
  \bibinfo{person}{Arvind Neelakantan}, \bibinfo{person}{Pranav Shyam},
  \bibinfo{person}{Girish Sastry}, \bibinfo{person}{Amanda Askell},
  \bibinfo{person}{Sandhini Agarwal}, \bibinfo{person}{Ariel Herbert{-}Voss},
  \bibinfo{person}{Gretchen Krueger}, \bibinfo{person}{Tom Henighan},
  \bibinfo{person}{Rewon Child}, \bibinfo{person}{Aditya Ramesh},
  \bibinfo{person}{Daniel~M. Ziegler}, \bibinfo{person}{Jeffrey Wu},
  \bibinfo{person}{Clemens Winter}, \bibinfo{person}{Christopher Hesse},
  \bibinfo{person}{Mark Chen}, \bibinfo{person}{Eric Sigler},
  \bibinfo{person}{Mateusz Litwin}, \bibinfo{person}{Scott Gray},
  \bibinfo{person}{Benjamin Chess}, \bibinfo{person}{Jack Clark},
  \bibinfo{person}{Christopher Berner}, \bibinfo{person}{Sam McCandlish},
  \bibinfo{person}{Alec Radford}, \bibinfo{person}{Ilya Sutskever}, {and}
  \bibinfo{person}{Dario Amodei}.} \bibinfo{year}{2020}\natexlab{}.
\newblock \showarticletitle{Language Models are Few-Shot Learners}. In
  \bibinfo{booktitle}{\emph{NeurIPS}}, Vol.~\bibinfo{volume}{33}.
  \bibinfo{pages}{1877--1901}.
\newblock


\bibitem[Cao et~al\mbox{.}(2023)]%
        {abs-2304-08191}
\bibfield{author}{\bibinfo{person}{Jialun Cao}, \bibinfo{person}{Meiziniu Li},
  \bibinfo{person}{Ming Wen}, {and} \bibinfo{person}{Shing{-}Chi Cheung}.}
  \bibinfo{year}{2023}\natexlab{}.
\newblock \showarticletitle{A study on Prompt Design, Advantages and
  Limitations of ChatGPT for Deep Learning Program Repair}.
\newblock \bibinfo{journal}{\emph{arXiv Preprint}}
  \bibinfo{volume}{https://arxiv.org/abs/2304.08191} (\bibinfo{year}{2023}).
\newblock


\bibitem[Cao et~al\mbox{.}(2022)]%
        {CaoSBWLT22}
\bibfield{author}{\bibinfo{person}{Sicong Cao}, \bibinfo{person}{Xiaobing Sun},
  \bibinfo{person}{Lili Bo}, \bibinfo{person}{Rongxin Wu}, \bibinfo{person}{Bin
  Li}, {and} \bibinfo{person}{Chuanqi Tao}.} \bibinfo{year}{2022}\natexlab{}.
\newblock \showarticletitle{{MVD:} Memory-Related Vulnerability Detection Based
  on Flow-Sensitive Graph Neural Networks}. In
  \bibinfo{booktitle}{\emph{{ICSE}}}. \bibinfo{pages}{1456--1468}.
\newblock


\bibitem[Chakraborty et~al\mbox{.}(2022)]%
        {ChakrabortyKDR22}
\bibfield{author}{\bibinfo{person}{Saikat Chakraborty}, \bibinfo{person}{Rahul
  Krishna}, \bibinfo{person}{Yangruibo Ding}, {and} \bibinfo{person}{Baishakhi
  Ray}.} \bibinfo{year}{2022}\natexlab{}.
\newblock \showarticletitle{Deep Learning Based Vulnerability Detection: Are We
  There Yet?}
\newblock \bibinfo{journal}{\emph{{IEEE} Trans. Software Eng.}}
  \bibinfo{volume}{48}, \bibinfo{number}{9} (\bibinfo{year}{2022}),
  \bibinfo{pages}{3280--3296}.
\newblock


\bibitem[Cheng et~al\mbox{.}(2022)]%
        {ChengZ0S22}
\bibfield{author}{\bibinfo{person}{Xiao Cheng}, \bibinfo{person}{Guanqin
  Zhang}, \bibinfo{person}{Haoyu Wang}, {and} \bibinfo{person}{Yulei Sui}.}
  \bibinfo{year}{2022}\natexlab{}.
\newblock \showarticletitle{Path-sensitive code embedding via contrastive
  learning for software vulnerability detection}. In
  \bibinfo{booktitle}{\emph{{ISSTA}}}. \bibinfo{pages}{519--531}.
\newblock


\bibitem[Choi et~al\mbox{.}(2017)]%
        {ChoiJOC17}
\bibfield{author}{\bibinfo{person}{Min{-}Je Choi}, \bibinfo{person}{Sehun
  Jeong}, \bibinfo{person}{Hakjoo Oh}, {and} \bibinfo{person}{Jaegul Choo}.}
  \bibinfo{year}{2017}\natexlab{}.
\newblock \showarticletitle{End-to-End Prediction of Buffer Overruns from Raw
  Source Code via Neural Memory Networks}. In
  \bibinfo{booktitle}{\emph{{IJCAI}}}. \bibinfo{pages}{1546--1553}.
\newblock


\bibitem[Chowdhery et~al\mbox{.}(2022)]%
        {abs-2204-02311}
\bibfield{author}{\bibinfo{person}{Aakanksha Chowdhery},
  \bibinfo{person}{Sharan Narang}, \bibinfo{person}{Jacob Devlin},
  \bibinfo{person}{Maarten Bosma}, \bibinfo{person}{Gaurav Mishra},
  \bibinfo{person}{Adam Roberts}, \bibinfo{person}{Paul Barham},
  \bibinfo{person}{Hyung~Won Chung}, \bibinfo{person}{Charles Sutton},
  \bibinfo{person}{Sebastian Gehrmann}, \bibinfo{person}{Parker Schuh},
  \bibinfo{person}{Kensen Shi}, \bibinfo{person}{Sasha Tsvyashchenko},
  \bibinfo{person}{Joshua Maynez}, \bibinfo{person}{Abhishek Rao},
  \bibinfo{person}{Parker Barnes}, \bibinfo{person}{Yi Tay},
  \bibinfo{person}{Noam Shazeer}, \bibinfo{person}{Vinodkumar Prabhakaran},
  \bibinfo{person}{Emily Reif}, \bibinfo{person}{Nan Du}, \bibinfo{person}{Ben
  Hutchinson}, \bibinfo{person}{Reiner Pope}, \bibinfo{person}{James Bradbury},
  \bibinfo{person}{Jacob Austin}, \bibinfo{person}{Michael Isard},
  \bibinfo{person}{Guy Gur{-}Ari}, \bibinfo{person}{Pengcheng Yin},
  \bibinfo{person}{Toju Duke}, \bibinfo{person}{Anselm Levskaya},
  \bibinfo{person}{Sanjay Ghemawat}, \bibinfo{person}{Sunipa Dev},
  \bibinfo{person}{Henryk Michalewski}, \bibinfo{person}{Xavier Garcia},
  \bibinfo{person}{Vedant Misra}, \bibinfo{person}{Kevin Robinson},
  \bibinfo{person}{Liam Fedus}, \bibinfo{person}{Denny Zhou},
  \bibinfo{person}{Daphne Ippolito}, \bibinfo{person}{David Luan},
  \bibinfo{person}{Hyeontaek Lim}, \bibinfo{person}{Barret Zoph},
  \bibinfo{person}{Alexander Spiridonov}, \bibinfo{person}{Ryan Sepassi},
  \bibinfo{person}{David Dohan}, \bibinfo{person}{Shivani Agrawal},
  \bibinfo{person}{Mark Omernick}, \bibinfo{person}{Andrew~M. Dai},
  \bibinfo{person}{Thanumalayan~Sankaranarayana Pillai}, \bibinfo{person}{Marie
  Pellat}, \bibinfo{person}{Aitor Lewkowycz}, \bibinfo{person}{Erica Moreira},
  \bibinfo{person}{Rewon Child}, \bibinfo{person}{Oleksandr Polozov},
  \bibinfo{person}{Katherine Lee}, \bibinfo{person}{Zongwei Zhou},
  \bibinfo{person}{Xuezhi Wang}, \bibinfo{person}{Brennan Saeta},
  \bibinfo{person}{Mark Diaz}, \bibinfo{person}{Orhan Firat},
  \bibinfo{person}{Michele Catasta}, \bibinfo{person}{Jason Wei},
  \bibinfo{person}{Kathy Meier{-}Hellstern}, \bibinfo{person}{Douglas Eck},
  \bibinfo{person}{Jeff Dean}, \bibinfo{person}{Slav Petrov}, {and}
  \bibinfo{person}{Noah Fiedel}.} \bibinfo{year}{2022}\natexlab{}.
\newblock \showarticletitle{PaLM: Scaling Language Modeling with Pathways}.
\newblock \bibinfo{journal}{\emph{arXiv Preprint}} (\bibinfo{year}{2022}).
\newblock
\urldef\tempurl%
\url{https://arxiv.org/abs/2204.02311}
\showURL{%
\tempurl}


\bibitem[Cui et~al\mbox{.}(2021)]%
        {CuiHJFY21}
\bibfield{author}{\bibinfo{person}{Lei Cui}, \bibinfo{person}{Zhiyu Hao},
  \bibinfo{person}{Yang Jiao}, \bibinfo{person}{Haiqiang Fei}, {and}
  \bibinfo{person}{Xiaochun Yun}.} \bibinfo{year}{2021}\natexlab{}.
\newblock \showarticletitle{VulDetector: Detecting Vulnerabilities Using
  Weighted Feature Graph Comparison}.
\newblock \bibinfo{journal}{\emph{{IEEE} Trans. Inf. Forensics Secur.}}
  \bibinfo{volume}{16} (\bibinfo{year}{2021}), \bibinfo{pages}{2004--2017}.
\newblock


\bibitem[Dam et~al\mbox{.}(2017)]%
        {abs-1708-02368}
\bibfield{author}{\bibinfo{person}{Hoa~Khanh Dam}, \bibinfo{person}{Truyen
  Tran}, \bibinfo{person}{Trang Pham}, \bibinfo{person}{Shien~Wee Ng},
  \bibinfo{person}{John Grundy}, {and} \bibinfo{person}{Aditya Ghose}.}
  \bibinfo{year}{2017}\natexlab{}.
\newblock \showarticletitle{Automatic feature learning for vulnerability
  prediction}.
\newblock \bibinfo{journal}{\emph{arXiv Preprint}} (\bibinfo{year}{2017}).
\newblock
\urldef\tempurl%
\url{https://arxiv.org/abs/1708.02368}
\showURL{%
\tempurl}


\bibitem[Diwan et~al\mbox{.}(2022)]%
        {DiwanLF22}
\bibfield{author}{\bibinfo{person}{Ashita Diwan}, \bibinfo{person}{Miles~Q.
  Li}, {and} \bibinfo{person}{Benjamin C.~M. Fung}.}
  \bibinfo{year}{2022}\natexlab{}.
\newblock \showarticletitle{VDGraph2Vec: Vulnerability Detection in Assembly
  Code using Message Passing Neural Networks}. In
  \bibinfo{booktitle}{\emph{{ICMLA}}}. \bibinfo{pages}{1039--1046}.
\newblock


\bibitem[Dong et~al\mbox{.}(2018)]%
        {DongWLXZ18}
\bibfield{author}{\bibinfo{person}{Feng Dong}, \bibinfo{person}{Junfeng Wang},
  \bibinfo{person}{Qi Li}, \bibinfo{person}{Guoai Xu}, {and}
  \bibinfo{person}{Shaodong Zhang}.} \bibinfo{year}{2018}\natexlab{}.
\newblock \showarticletitle{Defect Prediction in Android Binary Executables
  Using Deep Neural Network}.
\newblock \bibinfo{journal}{\emph{Wirel. Pers. Commun.}} \bibinfo{volume}{102},
  \bibinfo{number}{3} (\bibinfo{year}{2018}), \bibinfo{pages}{2261--2285}.
\newblock


\bibitem[Engler et~al\mbox{.}(2001)]%
        {EnglerCC01}
\bibfield{author}{\bibinfo{person}{Dawson~R. Engler}, \bibinfo{person}{David~Yu
  Chen}, {and} \bibinfo{person}{Andy Chou}.} \bibinfo{year}{2001}\natexlab{}.
\newblock \showarticletitle{Bugs as Deviant Behavior: {A} General Approach to
  Inferring Errors in Systems Code}. In \bibinfo{booktitle}{\emph{{SOSP}}}.
  \bibinfo{pages}{57--72}.
\newblock


\bibitem[Gao et~al\mbox{.}(2018)]%
        {GaoMSSZMMDDZC18}
\bibfield{author}{\bibinfo{person}{Qing Gao}, \bibinfo{person}{Sen Ma},
  \bibinfo{person}{Sihao Shao}, \bibinfo{person}{Yulei Sui},
  \bibinfo{person}{Guoliang Zhao}, \bibinfo{person}{Luyao Ma},
  \bibinfo{person}{Xiao Ma}, \bibinfo{person}{Fuyao Duan},
  \bibinfo{person}{Xiao Deng}, \bibinfo{person}{Shikun Zhang}, {and}
  \bibinfo{person}{Xianglong Chen}.} \bibinfo{year}{2018}\natexlab{}.
\newblock \showarticletitle{CoBOT: static {C/C++} bug detection in the presence
  of incomplete code}. In \bibinfo{booktitle}{\emph{{ICPC}}}.
  \bibinfo{pages}{385--388}.
\newblock


\bibitem[Gao et~al\mbox{.}(2021)]%
        {GaoFC20}
\bibfield{author}{\bibinfo{person}{Tianyu Gao}, \bibinfo{person}{Adam Fisch},
  {and} \bibinfo{person}{Danqi Chen}.} \bibinfo{year}{2021}\natexlab{}.
\newblock \showarticletitle{Making Pre-trained Language Models Better Few-shot
  Learners}. In \bibinfo{booktitle}{\emph{{ACL/IJCNLP} {(1)}}}.
  \bibinfo{pages}{3816--3830}.
\newblock


\bibitem[Grieco et~al\mbox{.}(2016)]%
        {GriecoGURFM16}
\bibfield{author}{\bibinfo{person}{Gustavo Grieco},
  \bibinfo{person}{Guillermo~Luis Grinblat}, \bibinfo{person}{Lucas~C. Uzal},
  \bibinfo{person}{Sanjay Rawat}, \bibinfo{person}{Josselin Feist}, {and}
  \bibinfo{person}{Laurent Mounier}.} \bibinfo{year}{2016}\natexlab{}.
\newblock \showarticletitle{Toward Large-Scale Vulnerability Discovery using
  Machine Learning}. In \bibinfo{booktitle}{\emph{{CODASPY}}}.
  \bibinfo{pages}{85--96}.
\newblock


\bibitem[Gu et~al\mbox{.}(2016)]%
        {GuZZK16}
\bibfield{author}{\bibinfo{person}{Xiaodong Gu}, \bibinfo{person}{Hongyu
  Zhang}, \bibinfo{person}{Dongmei Zhang}, {and} \bibinfo{person}{Sunghun
  Kim}.} \bibinfo{year}{2016}\natexlab{}.
\newblock \showarticletitle{Deep {API} learning}. In
  \bibinfo{booktitle}{\emph{{SIGSOFT} {FSE}}}. \bibinfo{pages}{631--642}.
\newblock


\bibitem[Gu et~al\mbox{.}(2022)]%
        {GuHLH22}
\bibfield{author}{\bibinfo{person}{Yuxian Gu}, \bibinfo{person}{Xu Han},
  \bibinfo{person}{Zhiyuan Liu}, {and} \bibinfo{person}{Minlie Huang}.}
  \bibinfo{year}{2022}\natexlab{}.
\newblock \showarticletitle{{PPT:} Pre-trained Prompt Tuning for Few-shot
  Learning}. In \bibinfo{booktitle}{\emph{{ACL} {(1)}}}.
  \bibinfo{pages}{8410--8423}.
\newblock


\bibitem[Guo et~al\mbox{.}(2021)]%
        {GuoRLFT0ZDSFTDC21}
\bibfield{author}{\bibinfo{person}{Daya Guo}, \bibinfo{person}{Shuo Ren},
  \bibinfo{person}{Shuai Lu}, \bibinfo{person}{Zhangyin Feng},
  \bibinfo{person}{Duyu Tang}, \bibinfo{person}{Shujie Liu},
  \bibinfo{person}{Long Zhou}, \bibinfo{person}{Nan Duan},
  \bibinfo{person}{Alexey Svyatkovskiy}, \bibinfo{person}{Shengyu Fu},
  \bibinfo{person}{Michele Tufano}, \bibinfo{person}{Shao~Kun Deng},
  \bibinfo{person}{Colin~B. Clement}, \bibinfo{person}{Dawn Drain},
  \bibinfo{person}{Neel Sundaresan}, \bibinfo{person}{Jian Yin},
  \bibinfo{person}{Daxin Jiang}, {and} \bibinfo{person}{Ming Zhou}.}
  \bibinfo{year}{2021}\natexlab{}.
\newblock \showarticletitle{GraphCodeBERT: Pre-training Code Representations
  with Data Flow}. In \bibinfo{booktitle}{\emph{{ICLR}}}.
\newblock
\urldef\tempurl%
\url{https://openreview.net/pdf?id=jLoC4ez43PZ}
\showURL{%
\tempurl}


\bibitem[Harer et~al\mbox{.}(2018)]%
        {abs-1803-04497}
\bibfield{author}{\bibinfo{person}{Jacob~A. Harer}, \bibinfo{person}{Louis~Y.
  Kim}, \bibinfo{person}{Rebecca~L. Russell}, \bibinfo{person}{Onur Ozdemir},
  \bibinfo{person}{Leonard~R. Kosta}, \bibinfo{person}{Akshay Rangamani},
  \bibinfo{person}{Lei~H. Hamilton}, \bibinfo{person}{Gabriel~I. Centeno},
  \bibinfo{person}{Jonathan~R. Key}, \bibinfo{person}{Paul~M. Ellingwood},
  \bibinfo{person}{Marc~W. McConley}, \bibinfo{person}{Jeffrey~M. Opper},
  \bibinfo{person}{Peter Chin}, {and} \bibinfo{person}{Tomo Lazovich}.}
  \bibinfo{year}{2018}\natexlab{}.
\newblock \showarticletitle{Automated software vulnerability detection with
  machine learning}.
\newblock \bibinfo{journal}{\emph{arXiv Preprint}} (\bibinfo{year}{2018}).
\newblock
\urldef\tempurl%
\url{https://arxiv.org/abs/1803.04497}
\showURL{%
\tempurl}


\bibitem[Hin et~al\mbox{.}(2022)]%
        {HinKCB22}
\bibfield{author}{\bibinfo{person}{David Hin}, \bibinfo{person}{Andrey Kan},
  \bibinfo{person}{Huaming Chen}, {and} \bibinfo{person}{Muhammad~Ali Babar}.}
  \bibinfo{year}{2022}\natexlab{}.
\newblock \showarticletitle{LineVD: Statement-level Vulnerability Detection
  using Graph Neural Networks}. In \bibinfo{booktitle}{\emph{{MSR}}}.
  \bibinfo{pages}{596--607}.
\newblock


\bibitem[Hoffmann et~al\mbox{.}(2022)]%
        {abs-2203-15556}
\bibfield{author}{\bibinfo{person}{Jordan Hoffmann}, \bibinfo{person}{Sebastian
  Borgeaud}, \bibinfo{person}{Arthur Mensch}, \bibinfo{person}{Elena
  Buchatskaya}, \bibinfo{person}{Trevor Cai}, \bibinfo{person}{Eliza
  Rutherford}, \bibinfo{person}{Diego de Las~Casas}, \bibinfo{person}{Lisa~Anne
  Hendricks}, \bibinfo{person}{Johannes Welbl}, \bibinfo{person}{Aidan Clark},
  \bibinfo{person}{Tom Hennigan}, \bibinfo{person}{Eric Noland},
  \bibinfo{person}{Katie Millican}, \bibinfo{person}{George van~den Driessche},
  \bibinfo{person}{Bogdan Damoc}, \bibinfo{person}{Aurelia Guy},
  \bibinfo{person}{Simon Osindero}, \bibinfo{person}{Karen Simonyan},
  \bibinfo{person}{Erich Elsen}, \bibinfo{person}{Jack~W. Rae},
  \bibinfo{person}{Oriol Vinyals}, {and} \bibinfo{person}{Laurent Sifre}.}
  \bibinfo{year}{2022}\natexlab{}.
\newblock \showarticletitle{Training Compute-Optimal Large Language Models}.
\newblock \bibinfo{journal}{\emph{arXiv Preprint}} (\bibinfo{year}{2022}).
\newblock
\urldef\tempurl%
\url{https://arxiv.org/abs/2203.15556}
\showURL{%
\tempurl}


\bibitem[Israel(2021)]%
        {Checkmarx}
\bibfield{author}{\bibinfo{person}{Israel}.} \bibinfo{year}{2021}\natexlab{}.
\newblock \bibinfo{booktitle}{\emph{Checkmarx}}.
\newblock
\urldef\tempurl%
\url{https://www.checkmarx.com/}
\showURL{%
\tempurl}


\bibitem[Jiang et~al\mbox{.}(2020)]%
        {JiangXAN20}
\bibfield{author}{\bibinfo{person}{Zhengbao Jiang}, \bibinfo{person}{Frank~F.
  Xu}, \bibinfo{person}{Jun Araki}, {and} \bibinfo{person}{Graham Neubig}.}
  \bibinfo{year}{2020}\natexlab{}.
\newblock \showarticletitle{How Can We Know What Language Models Know}.
\newblock \bibinfo{journal}{\emph{Trans. Assoc. Comput. Linguistics}}
  \bibinfo{volume}{8} (\bibinfo{year}{2020}), \bibinfo{pages}{423--438}.
\newblock


\bibitem[Le et~al\mbox{.}(2019)]%
        {LeNLPMVQ19}
\bibfield{author}{\bibinfo{person}{Tue Le}, \bibinfo{person}{Tuan Nguyen},
  \bibinfo{person}{Trung Le}, \bibinfo{person}{Dinh~Q. Phung},
  \bibinfo{person}{Paul Montague}, \bibinfo{person}{Olivier~Y. de Vel}, {and}
  \bibinfo{person}{Lizhen Qu}.} \bibinfo{year}{2019}\natexlab{}.
\newblock \showarticletitle{Maximal Divergence Sequential Autoencoder for
  Binary Software Vulnerability Detection}. In
  \bibinfo{booktitle}{\emph{{ICLR}}}.
\newblock
\urldef\tempurl%
\url{https://openreview.net/pdf?id=ByloIiCqYQ}
\showURL{%
\tempurl}


\bibitem[Lester et~al\mbox{.}(2021)]%
        {LesterAC21}
\bibfield{author}{\bibinfo{person}{Brian Lester}, \bibinfo{person}{Rami
  Al{-}Rfou}, {and} \bibinfo{person}{Noah Constant}.}
  \bibinfo{year}{2021}\natexlab{}.
\newblock \showarticletitle{The Power of Scale for Parameter-Efficient Prompt
  Tuning}. In \bibinfo{booktitle}{\emph{{EMNLP} {(1)}}}.
  \bibinfo{pages}{3045--3059}.
\newblock


\bibitem[Li and Liang(2021)]%
        {LiL20}
\bibfield{author}{\bibinfo{person}{Xiang~Lisa Li} {and} \bibinfo{person}{Percy
  Liang}.} \bibinfo{year}{2021}\natexlab{}.
\newblock \showarticletitle{Prefix-Tuning: Optimizing Continuous Prompts for
  Generation}. In \bibinfo{booktitle}{\emph{{ACL/IJCNLP} {(1)}}}.
  \bibinfo{pages}{4582--4597}.
\newblock


\bibitem[Li and Zhou(2005)]%
        {LiZ05}
\bibfield{author}{\bibinfo{person}{Zhenmin Li} {and} \bibinfo{person}{Yuanyuan
  Zhou}.} \bibinfo{year}{2005}\natexlab{}.
\newblock \showarticletitle{PR-Miner: automatically extracting implicit
  programming rules and detecting violations in large software code}. In
  \bibinfo{booktitle}{\emph{{ESEC/SIGSOFT} {FSE}}}. \bibinfo{pages}{306--315}.
\newblock


\bibitem[Li et~al\mbox{.}(2022)]%
        {LiZXJZC22}
\bibfield{author}{\bibinfo{person}{Zhen Li}, \bibinfo{person}{Deqing Zou},
  \bibinfo{person}{Shouhuai Xu}, \bibinfo{person}{Hai Jin},
  \bibinfo{person}{Yawei Zhu}, {and} \bibinfo{person}{Zhaoxuan Chen}.}
  \bibinfo{year}{2022}\natexlab{}.
\newblock \showarticletitle{SySeVR: {A} Framework for Using Deep Learning to
  Detect Software Vulnerabilities}.
\newblock \bibinfo{journal}{\emph{{IEEE} Trans. Dependable Secur. Comput.}}
  \bibinfo{volume}{19}, \bibinfo{number}{4} (\bibinfo{year}{2022}),
  \bibinfo{pages}{2244--2258}.
\newblock


\bibitem[Li et~al\mbox{.}(2018)]%
        {LiZXO0WDZ18}
\bibfield{author}{\bibinfo{person}{Zhen Li}, \bibinfo{person}{Deqing Zou},
  \bibinfo{person}{Shouhuai Xu}, \bibinfo{person}{Xinyu Ou},
  \bibinfo{person}{Hai Jin}, \bibinfo{person}{Sujuan Wang},
  \bibinfo{person}{Zhijun Deng}, {and} \bibinfo{person}{Yuyi Zhong}.}
  \bibinfo{year}{2018}\natexlab{}.
\newblock \showarticletitle{VulDeePecker: {A} Deep Learning-Based System for
  Vulnerability Detection}. In \bibinfo{booktitle}{\emph{{NDSS}}}.
\newblock


\bibitem[Lin et~al\mbox{.}(2020)]%
        {LinWHZX20}
\bibfield{author}{\bibinfo{person}{Guanjun Lin}, \bibinfo{person}{Sheng Wen},
  \bibinfo{person}{Qing{-}Long Han}, \bibinfo{person}{Jun Zhang}, {and}
  \bibinfo{person}{Yang Xiang}.} \bibinfo{year}{2020}\natexlab{}.
\newblock \showarticletitle{Software Vulnerability Detection Using Deep Neural
  Networks: {A} Survey}.
\newblock \bibinfo{journal}{\emph{Proc. {IEEE}}} \bibinfo{volume}{108},
  \bibinfo{number}{10} (\bibinfo{year}{2020}), \bibinfo{pages}{1825--1848}.
\newblock


\bibitem[Lin et~al\mbox{.}(2021)]%
        {LinZLPVMX21}
\bibfield{author}{\bibinfo{person}{Guanjun Lin}, \bibinfo{person}{Jun Zhang},
  \bibinfo{person}{Wei Luo}, \bibinfo{person}{Lei Pan},
  \bibinfo{person}{Olivier~Y. de Vel}, \bibinfo{person}{Paul Montague}, {and}
  \bibinfo{person}{Yang Xiang}.} \bibinfo{year}{2021}\natexlab{}.
\newblock \showarticletitle{Software Vulnerability Discovery via Learning
  Multi-Domain Knowledge Bases}.
\newblock \bibinfo{journal}{\emph{{IEEE} Trans. Dependable Secur. Comput.}}
  \bibinfo{volume}{18}, \bibinfo{number}{5} (\bibinfo{year}{2021}),
  \bibinfo{pages}{2469--2485}.
\newblock


\bibitem[Lin et~al\mbox{.}(2017)]%
        {LinZLPX17}
\bibfield{author}{\bibinfo{person}{Guanjun Lin}, \bibinfo{person}{Jun Zhang},
  \bibinfo{person}{Wei Luo}, \bibinfo{person}{Lei Pan}, {and}
  \bibinfo{person}{Yang Xiang}.} \bibinfo{year}{2017}\natexlab{}.
\newblock \showarticletitle{{POSTER:} Vulnerability Discovery with Function
  Representation Learning from Unlabeled Projects}. In
  \bibinfo{booktitle}{\emph{{CCS}}}. \bibinfo{pages}{2539--2541}.
\newblock


\bibitem[Lin et~al\mbox{.}(2018)]%
        {LinZLPXVM18}
\bibfield{author}{\bibinfo{person}{Guanjun Lin}, \bibinfo{person}{Jun Zhang},
  \bibinfo{person}{Wei Luo}, \bibinfo{person}{Lei Pan}, \bibinfo{person}{Yang
  Xiang}, \bibinfo{person}{Olivier~Y. de Vel}, {and} \bibinfo{person}{Paul
  Montague}.} \bibinfo{year}{2018}\natexlab{}.
\newblock \showarticletitle{Cross-Project Transfer Representation Learning for
  Vulnerable Function Discovery}.
\newblock \bibinfo{journal}{\emph{{IEEE} Trans. Ind. Informatics}}
  \bibinfo{volume}{14}, \bibinfo{number}{7} (\bibinfo{year}{2018}),
  \bibinfo{pages}{3289--3297}.
\newblock


\bibitem[Lipp et~al\mbox{.}(2022)]%
        {LippBP22}
\bibfield{author}{\bibinfo{person}{Stephan Lipp}, \bibinfo{person}{Sebastian
  Banescu}, {and} \bibinfo{person}{Alexander Pretschner}.}
  \bibinfo{year}{2022}\natexlab{}.
\newblock \showarticletitle{An empirical study on the effectiveness of static
  {C} code analyzers for vulnerability detection}. In
  \bibinfo{booktitle}{\emph{{ISSTA}}}. \bibinfo{pages}{544--555}.
\newblock


\bibitem[Liu et~al\mbox{.}(2023a)]%
        {abs-2305-08360}
\bibfield{author}{\bibinfo{person}{Chao Liu}, \bibinfo{person}{Xuanlin Bao},
  \bibinfo{person}{Hongyu Zhang}, \bibinfo{person}{Neng Zhang},
  \bibinfo{person}{Haibo Hu}, \bibinfo{person}{Xiaohong Zhang}, {and}
  \bibinfo{person}{Meng Yan}.} \bibinfo{year}{2023}\natexlab{a}.
\newblock \showarticletitle{Improving ChatGPT Prompt for Code Generation}.
\newblock \bibinfo{journal}{\emph{arXiv Preprint}} (\bibinfo{year}{2023}).
\newblock
\urldef\tempurl%
\url{https://arxiv.org/abs/2305.08360}
\showURL{%
\tempurl}


\bibitem[Liu et~al\mbox{.}(2022)]%
        {LiuSZDCC22}
\bibfield{author}{\bibinfo{person}{Jiachang Liu}, \bibinfo{person}{Dinghan
  Shen}, \bibinfo{person}{Yizhe Zhang}, \bibinfo{person}{Bill Dolan},
  \bibinfo{person}{Lawrence Carin}, {and} \bibinfo{person}{Weizhu Chen}.}
  \bibinfo{year}{2022}\natexlab{}.
\newblock \showarticletitle{What Makes Good In-Context Examples for GPT-3?}. In
  \bibinfo{booktitle}{\emph{DeeLIO@ACL}}. \bibinfo{pages}{100--114}.
\newblock


\bibitem[Liu et~al\mbox{.}(2023b)]%
        {LiuYFJHN23}
\bibfield{author}{\bibinfo{person}{Pengfei Liu}, \bibinfo{person}{Weizhe Yuan},
  \bibinfo{person}{Jinlan Fu}, \bibinfo{person}{Zhengbao Jiang},
  \bibinfo{person}{Hiroaki Hayashi}, {and} \bibinfo{person}{Graham Neubig}.}
  \bibinfo{year}{2023}\natexlab{b}.
\newblock \showarticletitle{Pre-train, Prompt, and Predict: {A} Systematic
  Survey of Prompting Methods in Natural Language Processing}.
\newblock \bibinfo{journal}{\emph{{ACM} Comput. Surv.}} \bibinfo{volume}{55},
  \bibinfo{number}{9} (\bibinfo{year}{2023}), \bibinfo{pages}{195:1--195:35}.
\newblock


\bibitem[Liu and Chilton(2022)]%
        {LiuC22a}
\bibfield{author}{\bibinfo{person}{Vivian Liu} {and} \bibinfo{person}{Lydia~B.
  Chilton}.} \bibinfo{year}{2022}\natexlab{}.
\newblock \showarticletitle{Design Guidelines for Prompt Engineering
  Text-to-Image Generative Models}. In \bibinfo{booktitle}{\emph{{CHI}}}.
  \bibinfo{pages}{384:1--384:23}.
\newblock


\bibitem[Liu et~al\mbox{.}(2021)]%
        {abs-2103-10385}
\bibfield{author}{\bibinfo{person}{Xiao Liu}, \bibinfo{person}{Yanan Zheng},
  \bibinfo{person}{Zhengxiao Du}, \bibinfo{person}{Ming Ding},
  \bibinfo{person}{Yujie Qian}, \bibinfo{person}{Zhilin Yang}, {and}
  \bibinfo{person}{Jie Tang}.} \bibinfo{year}{2021}\natexlab{}.
\newblock \showarticletitle{{GPT} Understands, Too}.
\newblock \bibinfo{journal}{\emph{arXiv Preprint}} (\bibinfo{year}{2021}).
\newblock
\urldef\tempurl%
\url{https://arxiv.org/abs/2103.10385}
\showURL{%
\tempurl}


\bibitem[Luo et~al\mbox{.}(2023)]%
        {abs-2306-08568}
\bibfield{author}{\bibinfo{person}{Ziyang Luo}, \bibinfo{person}{Can Xu},
  \bibinfo{person}{Pu Zhao}, \bibinfo{person}{Qingfeng Sun},
  \bibinfo{person}{Xiubo Geng}, \bibinfo{person}{Wenxiang Hu},
  \bibinfo{person}{Chongyang Tao}, \bibinfo{person}{Jing Ma},
  \bibinfo{person}{Qingwei Lin}, {and} \bibinfo{person}{Daxin Jiang}.}
  \bibinfo{year}{2023}\natexlab{}.
\newblock \showarticletitle{WizardCoder: Empowering Code Large Language Models
  with Evol-Instruct}.
\newblock \bibinfo{journal}{\emph{arXiv Preprint}} (\bibinfo{year}{2023}).
\newblock
\urldef\tempurl%
\url{https://arxiv.org/abs/2306.08568}
\showURL{%
\tempurl}


\bibitem[Maddigan and Susnjak(2023)]%
        {abs-2302-02094}
\bibfield{author}{\bibinfo{person}{Paula Maddigan} {and} \bibinfo{person}{Teo
  Susnjak}.} \bibinfo{year}{2023}\natexlab{}.
\newblock \showarticletitle{Chat2VIS: Generating Data Visualisations via
  Natural Language using ChatGPT, Codex and {GPT-3} Large Language Models}.
\newblock \bibinfo{journal}{\emph{arXiv Preprint}} (\bibinfo{year}{2023}).
\newblock
\urldef\tempurl%
\url{https://arxiv.org/abs/2302.02094}
\showURL{%
\tempurl}


\bibitem[Mikolov et~al\mbox{.}(2013)]%
        {abs-1301-3781}
\bibfield{author}{\bibinfo{person}{Tom{\'{a}}s Mikolov}, \bibinfo{person}{Kai
  Chen}, \bibinfo{person}{Greg Corrado}, {and} \bibinfo{person}{Jeffrey Dean}.}
  \bibinfo{year}{2013}\natexlab{}.
\newblock \showarticletitle{Efficient Estimation of Word Representations in
  Vector Space}. In \bibinfo{booktitle}{\emph{{ICLR}}}.
\newblock
\urldef\tempurl%
\url{https://arxiv.org/abs/1301.3781}
\showURL{%
\tempurl}


\bibitem[Mon et~al\mbox{.}(2023)]%
        {MonKCM23}
\bibfield{author}{\bibinfo{person}{Khine~Yin Mon}, \bibinfo{person}{Masanari
  Kondo}, \bibinfo{person}{Eunjong Choi}, {and} \bibinfo{person}{Osamu
  Mizuno}.} \bibinfo{year}{2023}\natexlab{}.
\newblock \showarticletitle{Commit-Based Class-Level Defect Prediction for
  Python Projects}.
\newblock \bibinfo{journal}{\emph{{IEICE} Trans. Inf. Syst.}}
  \bibinfo{volume}{106}, \bibinfo{number}{2} (\bibinfo{year}{2023}),
  \bibinfo{pages}{157--165}.
\newblock


\bibitem[Nguyen et~al\mbox{.}(2022)]%
        {NguyenNNLTP22}
\bibfield{author}{\bibinfo{person}{Van{-}Anh Nguyen}, \bibinfo{person}{Dai~Quoc
  Nguyen}, \bibinfo{person}{Van Nguyen}, \bibinfo{person}{Trung Le},
  \bibinfo{person}{Quan~Hung Tran}, {and} \bibinfo{person}{Dinh Phung}.}
  \bibinfo{year}{2022}\natexlab{}.
\newblock \showarticletitle{ReGVD: Revisiting Graph Neural Networks for
  Vulnerability Detection}. In \bibinfo{booktitle}{\emph{ICSE-Companion}}.
  \bibinfo{pages}{178--182}.
\newblock


\bibitem[OpenAI(2023a)]%
        {abs-2303-08774}
\bibfield{author}{\bibinfo{person}{OpenAI}.} \bibinfo{year}{2023}\natexlab{a}.
\newblock \showarticletitle{{GPT-4} Technical Report}.
\newblock \bibinfo{journal}{\emph{arXiv Preprint}} (\bibinfo{year}{2023}).
\newblock
\urldef\tempurl%
\url{https://arxiv.org/abs/2303.08774}
\showURL{%
\tempurl}


\bibitem[OpenAI(2023b)]%
        {GBP}
\bibfield{author}{\bibinfo{person}{OpenAI}.} \bibinfo{year}{2023}\natexlab{b}.
\newblock \bibinfo{booktitle}{\emph{gpt-best-practices}}.
\newblock
\urldef\tempurl%
\url{https://platform.openai.com/docs/guides/gpt-best-practices}
\showURL{%
\tempurl}


\bibitem[Pang et~al\mbox{.}(2015)]%
        {PangXN15}
\bibfield{author}{\bibinfo{person}{Yulei Pang}, \bibinfo{person}{Xiaozhen Xue},
  {and} \bibinfo{person}{Akbar~Siami Namin}.} \bibinfo{year}{2015}\natexlab{}.
\newblock \showarticletitle{Predicting Vulnerable Software Components through
  N-Gram Analysis and Statistical Feature Selection}. In
  \bibinfo{booktitle}{\emph{{ICMLA}}}. \bibinfo{pages}{543--548}.
\newblock


\bibitem[Peng et~al\mbox{.}(2015)]%
        {PengMLLZJ15}
\bibfield{author}{\bibinfo{person}{Hao Peng}, \bibinfo{person}{Lili Mou},
  \bibinfo{person}{Ge Li}, \bibinfo{person}{Yuxuan Liu}, \bibinfo{person}{Lu
  Zhang}, {and} \bibinfo{person}{Zhi Jin}.} \bibinfo{year}{2015}\natexlab{}.
\newblock \showarticletitle{Building Program Vector Representations for Deep
  Learning}. In \bibinfo{booktitle}{\emph{{KSEM}}},
  Vol.~\bibinfo{volume}{9403}. \bibinfo{pages}{547--553}.
\newblock


\bibitem[Pitis et~al\mbox{.}(2023)]%
        {abs-2304-05970}
\bibfield{author}{\bibinfo{person}{Silviu Pitis}, \bibinfo{person}{Michael~R.
  Zhang}, \bibinfo{person}{Andrew Wang}, {and} \bibinfo{person}{Jimmy Ba}.}
  \bibinfo{year}{2023}\natexlab{}.
\newblock \showarticletitle{Boosted Prompt Ensembles for Large Language
  Models}.
\newblock \bibinfo{journal}{\emph{arXiv Preprint}} (\bibinfo{year}{2023}).
\newblock
\urldef\tempurl%
\url{https://arxiv.org/abs/2304.05970}
\showURL{%
\tempurl}


\bibitem[Qin and Eisner(2021)]%
        {QinE21}
\bibfield{author}{\bibinfo{person}{Guanghui Qin} {and} \bibinfo{person}{Jason
  Eisner}.} \bibinfo{year}{2021}\natexlab{}.
\newblock \showarticletitle{Learning How to Ask: Querying LMs with Mixtures of
  Soft Prompts}. In \bibinfo{booktitle}{\emph{{NAACL-HLT}}}.
  \bibinfo{pages}{5203--5212}.
\newblock


\bibitem[Russell et~al\mbox{.}(2018)]%
        {RussellKHLHOEM18}
\bibfield{author}{\bibinfo{person}{Rebecca~L. Russell},
  \bibinfo{person}{Louis~Y. Kim}, \bibinfo{person}{Lei~H. Hamilton},
  \bibinfo{person}{Tomo Lazovich}, \bibinfo{person}{Jacob Harer},
  \bibinfo{person}{Onur Ozdemir}, \bibinfo{person}{Paul~M. Ellingwood}, {and}
  \bibinfo{person}{Marc~W. McConley}.} \bibinfo{year}{2018}\natexlab{}.
\newblock \showarticletitle{Automated Vulnerability Detection in Source Code
  Using Deep Representation Learning}. In \bibinfo{booktitle}{\emph{{ICMLA}}}.
  \bibinfo{pages}{757--762}.
\newblock


\bibitem[Scandariato et~al\mbox{.}(2014)]%
        {ScandariatoWHJ14}
\bibfield{author}{\bibinfo{person}{Riccardo Scandariato},
  \bibinfo{person}{James Walden}, \bibinfo{person}{Aram Hovsepyan}, {and}
  \bibinfo{person}{Wouter Joosen}.} \bibinfo{year}{2014}\natexlab{}.
\newblock \showarticletitle{Predicting Vulnerable Software Components via Text
  Mining}.
\newblock \bibinfo{journal}{\emph{{IEEE} Trans. Software Eng.}}
  \bibinfo{volume}{40}, \bibinfo{number}{10} (\bibinfo{year}{2014}),
  \bibinfo{pages}{993--1006}.
\newblock


\bibitem[Sestili et~al\mbox{.}(2018)]%
        {abs-1808-09897}
\bibfield{author}{\bibinfo{person}{Carson~D. Sestili},
  \bibinfo{person}{William~S. Snavely}, {and} \bibinfo{person}{Nathan~M.
  VanHoudnos}.} \bibinfo{year}{2018}\natexlab{}.
\newblock \showarticletitle{Towards security defect prediction with {AI}}.
\newblock \bibinfo{journal}{\emph{arXiv Preprint}} (\bibinfo{year}{2018}).
\newblock
\urldef\tempurl%
\url{https://arxiv.org/abs/1808.09897}
\showURL{%
\tempurl}


\bibitem[Shar et~al\mbox{.}(2015)]%
        {SharBT15}
\bibfield{author}{\bibinfo{person}{Lwin~Khin Shar}, \bibinfo{person}{Lionel~C.
  Briand}, {and} \bibinfo{person}{Hee Beng~Kuan Tan}.}
  \bibinfo{year}{2015}\natexlab{}.
\newblock \showarticletitle{Web Application Vulnerability Prediction Using
  Hybrid Program Analysis and Machine Learning}.
\newblock \bibinfo{journal}{\emph{{IEEE} Trans. Dependable Secur. Comput.}}
  \bibinfo{volume}{12}, \bibinfo{number}{6} (\bibinfo{year}{2015}),
  \bibinfo{pages}{688--707}.
\newblock


\bibitem[Shar and Tan(2013)]%
        {SharT13}
\bibfield{author}{\bibinfo{person}{Lwin~Khin Shar} {and} \bibinfo{person}{Hee
  Beng~Kuan Tan}.} \bibinfo{year}{2013}\natexlab{}.
\newblock \showarticletitle{Predicting {SQL} injection and cross site scripting
  vulnerabilities through mining input sanitization patterns}.
\newblock \bibinfo{journal}{\emph{Inf. Softw. Technol.}} \bibinfo{volume}{55},
  \bibinfo{number}{10} (\bibinfo{year}{2013}), \bibinfo{pages}{1767--1780}.
\newblock


\bibitem[Shin et~al\mbox{.}(2020)]%
        {ShinRLWS20}
\bibfield{author}{\bibinfo{person}{Taylor Shin}, \bibinfo{person}{Yasaman
  Razeghi}, \bibinfo{person}{Robert L.~Logan IV}, \bibinfo{person}{Eric
  Wallace}, {and} \bibinfo{person}{Sameer Singh}.}
  \bibinfo{year}{2020}\natexlab{}.
\newblock \showarticletitle{AutoPrompt: Eliciting Knowledge from Language
  Models with Automatically Generated Prompts}. In
  \bibinfo{booktitle}{\emph{{EMNLP} {(1)}}}. \bibinfo{pages}{4222--4235}.
\newblock


\bibitem[Sobania et~al\mbox{.}(2023)]%
        {abs-2301-08653}
\bibfield{author}{\bibinfo{person}{Dominik Sobania}, \bibinfo{person}{Martin
  Briesch}, \bibinfo{person}{Carol Hanna}, {and} \bibinfo{person}{Justyna
  Petke}.} \bibinfo{year}{2023}\natexlab{}.
\newblock \showarticletitle{An Analysis of the Automatic Bug Fixing Performance
  of ChatGPT}.
\newblock \bibinfo{journal}{\emph{arXiv Preprint}} (\bibinfo{year}{2023}).
\newblock
\urldef\tempurl%
\url{https://arxiv.org/abs/2301.08653}
\showURL{%
\tempurl}


\bibitem[Srivastava et~al\mbox{.}(2022)]%
        {abs-2206-04615}
\bibfield{author}{\bibinfo{person}{Aarohi Srivastava}, \bibinfo{person}{Abhinav
  Rastogi}, \bibinfo{person}{Abhishek Rao}, \bibinfo{person}{Abu Awal~Md
  Shoeb}, \bibinfo{person}{Abubakar Abid}, \bibinfo{person}{Adam Fisch},
  \bibinfo{person}{Adam~R. Brown}, \bibinfo{person}{Adam Santoro},
  \bibinfo{person}{Aditya Gupta}, \bibinfo{person}{Adri{\`{a}}
  Garriga{-}Alonso}, \bibinfo{person}{Agnieszka Kluska}, \bibinfo{person}{Aitor
  Lewkowycz}, \bibinfo{person}{Akshat Agarwal}, \bibinfo{person}{Alethea
  Power}, \bibinfo{person}{Alex Ray}, \bibinfo{person}{Alex Warstadt},
  \bibinfo{person}{Alexander~W. Kocurek}, \bibinfo{person}{Ali Safaya},
  \bibinfo{person}{Ali Tazarv}, \bibinfo{person}{Alice Xiang},
  \bibinfo{person}{Alicia Parrish}, \bibinfo{person}{Allen Nie},
  \bibinfo{person}{Aman Hussain}, \bibinfo{person}{Amanda Askell},
  \bibinfo{person}{Amanda Dsouza}, \bibinfo{person}{Ameet Rahane},
  \bibinfo{person}{Anantharaman~S. Iyer}, \bibinfo{person}{Anders Andreassen},
  \bibinfo{person}{Andrea Santilli}, \bibinfo{person}{Andreas
  Stuhlm{\"{u}}ller}, \bibinfo{person}{Andrew~M. Dai}, \bibinfo{person}{Andrew
  La}, \bibinfo{person}{Andrew~K. Lampinen}, \bibinfo{person}{Andy Zou},
  \bibinfo{person}{Angela Jiang}, \bibinfo{person}{Angelica Chen},
  \bibinfo{person}{Anh Vuong}, \bibinfo{person}{Animesh Gupta},
  \bibinfo{person}{Anna Gottardi}, \bibinfo{person}{Antonio Norelli},
  \bibinfo{person}{Anu Venkatesh}, \bibinfo{person}{Arash Gholamidavoodi},
  \bibinfo{person}{Arfa Tabassum}, \bibinfo{person}{Arul Menezes},
  \bibinfo{person}{Arun Kirubarajan}, \bibinfo{person}{Asher Mullokandov},
  \bibinfo{person}{Ashish Sabharwal}, \bibinfo{person}{Austin Herrick},
  \bibinfo{person}{Avia Efrat}, \bibinfo{person}{Aykut Erdem},
  \bibinfo{person}{Ayla Karakas}, {and} \bibinfo{person}{et al.}}
  \bibinfo{year}{2022}\natexlab{}.
\newblock \showarticletitle{Beyond the Imitation Game: Quantifying and
  extrapolating the capabilities of language models}.
\newblock \bibinfo{journal}{\emph{arXiv Preprint}} (\bibinfo{year}{2022}).
\newblock
\urldef\tempurl%
\url{https://arxiv.org/abs/2206.04615}
\showURL{%
\tempurl}


\bibitem[Sui and Xue(2016)]%
        {SuiX16}
\bibfield{author}{\bibinfo{person}{Yulei Sui} {and} \bibinfo{person}{Jingling
  Xue}.} \bibinfo{year}{2016}\natexlab{}.
\newblock \showarticletitle{{SVF:} interprocedural static value-flow analysis
  in {LLVM}}. In \bibinfo{booktitle}{\emph{{CC}}}. \bibinfo{pages}{265--266}.
\newblock


\bibitem[Tay et~al\mbox{.}({[n.\,d.]})]%
        {Tay00GW0CBSZZHM23}
\bibfield{author}{\bibinfo{person}{Yi Tay}, \bibinfo{person}{Mostafa Dehghani},
  \bibinfo{person}{Vinh~Q. Tran}, \bibinfo{person}{Xavier Garcia},
  \bibinfo{person}{Jason Wei}, \bibinfo{person}{Xuezhi Wang},
  \bibinfo{person}{Hyung~Won Chung}, \bibinfo{person}{Dara Bahri},
  \bibinfo{person}{Tal Schuster}, \bibinfo{person}{Huaixiu~Steven Zheng},
  \bibinfo{person}{Denny Zhou}, \bibinfo{person}{Neil Houlsby}, {and}
  \bibinfo{person}{Donald Metzler}.} \bibinfo{year}{[n.\,d.]}\natexlab{}.
\newblock \showarticletitle{{UL2:} Unifying Language Learning Paradigms}. In
  \bibinfo{booktitle}{\emph{{ICLR}}}.
\newblock
\urldef\tempurl%
\url{https://openreview.net/pdf?id=6ruVLB727MC}
\showURL{%
\tempurl}


\bibitem[Touvron et~al\mbox{.}(2023)]%
        {abs-2302-13971}
\bibfield{author}{\bibinfo{person}{Hugo Touvron}, \bibinfo{person}{Thibaut
  Lavril}, \bibinfo{person}{Gautier Izacard}, \bibinfo{person}{Xavier
  Martinet}, \bibinfo{person}{Marie{-}Anne Lachaux},
  \bibinfo{person}{Timoth{\'{e}}e Lacroix}, \bibinfo{person}{Baptiste
  Rozi{\`{e}}re}, \bibinfo{person}{Naman Goyal}, \bibinfo{person}{Eric Hambro},
  \bibinfo{person}{Faisal Azhar}, \bibinfo{person}{Aur{\'{e}}lien Rodriguez},
  \bibinfo{person}{Armand Joulin}, \bibinfo{person}{Edouard Grave}, {and}
  \bibinfo{person}{Guillaume Lample}.} \bibinfo{year}{2023}\natexlab{}.
\newblock \showarticletitle{LLaMA: Open and Efficient Foundation Language
  Models}.
\newblock \bibinfo{journal}{\emph{arXiv Preprint}} (\bibinfo{year}{2023}).
\newblock
\urldef\tempurl%
\url{https://arxiv.org/abs/2302.13971}
\showURL{%
\tempurl}


\bibitem[Wang et~al\mbox{.}(2021)]%
        {WangYTTHFFBW21}
\bibfield{author}{\bibinfo{person}{Huanting Wang}, \bibinfo{person}{Guixin Ye},
  \bibinfo{person}{Zhanyong Tang}, \bibinfo{person}{Shin~Hwei Tan},
  \bibinfo{person}{Songfang Huang}, \bibinfo{person}{Dingyi Fang},
  \bibinfo{person}{Yansong Feng}, \bibinfo{person}{Lizhong Bian}, {and}
  \bibinfo{person}{Zheng Wang}.} \bibinfo{year}{2021}\natexlab{}.
\newblock \showarticletitle{Combining Graph-Based Learning With Automated Data
  Collection for Code Vulnerability Detection}.
\newblock \bibinfo{journal}{\emph{{IEEE} Trans. Inf. Forensics Secur.}}
  \bibinfo{volume}{16} (\bibinfo{year}{2021}), \bibinfo{pages}{1943--1958}.
\newblock


\bibitem[Wang et~al\mbox{.}(2016a)]%
        {WangCMT16}
\bibfield{author}{\bibinfo{person}{Song Wang}, \bibinfo{person}{Devin Chollak},
  \bibinfo{person}{Dana Movshovitz{-}Attias}, {and} \bibinfo{person}{Lin Tan}.}
  \bibinfo{year}{2016}\natexlab{a}.
\newblock \showarticletitle{Bugram: bug detection with n-gram language models}.
  In \bibinfo{booktitle}{\emph{{ASE}}}. \bibinfo{pages}{708--719}.
\newblock


\bibitem[Wang et~al\mbox{.}(2016b)]%
        {WangLT16}
\bibfield{author}{\bibinfo{person}{Song Wang}, \bibinfo{person}{Taiyue Liu},
  {and} \bibinfo{person}{Lin Tan}.} \bibinfo{year}{2016}\natexlab{b}.
\newblock \showarticletitle{Automatically learning semantic features for defect
  prediction}. In \bibinfo{booktitle}{\emph{{ICSE}}}.
  \bibinfo{pages}{297--308}.
\newblock


\bibitem[Wei et~al\mbox{.}(2022)]%
        {Wei0SBIXCLZ22}
\bibfield{author}{\bibinfo{person}{Jason Wei}, \bibinfo{person}{Xuezhi Wang},
  \bibinfo{person}{Dale Schuurmans}, \bibinfo{person}{Maarten Bosma},
  \bibinfo{person}{Brian Ichter}, \bibinfo{person}{Fei Xia},
  \bibinfo{person}{Ed~H. Chi}, \bibinfo{person}{Quoc~V. Le}, {and}
  \bibinfo{person}{Denny Zhou}.} \bibinfo{year}{2022}\natexlab{}.
\newblock \showarticletitle{Chain-of-Thought Prompting Elicits Reasoning in
  Large Language Models}. In \bibinfo{booktitle}{\emph{NeurIPS}},
  Vol.~\bibinfo{volume}{35}. \bibinfo{pages}{24824--24837}.
\newblock


\bibitem[Wen et~al\mbox{.}(2015)]%
        {WenHCXZJ15}
\bibfield{author}{\bibinfo{person}{Sheng Wen}, \bibinfo{person}{Mohammad~Sayad
  Haghighi}, \bibinfo{person}{Chao Chen}, \bibinfo{person}{Yang Xiang},
  \bibinfo{person}{Wanlei Zhou}, {and} \bibinfo{person}{Weijia Jia}.}
  \bibinfo{year}{2015}\natexlab{}.
\newblock \showarticletitle{A Sword with Two Edges: Propagation Studies on Both
  Positive and Negative Information in Online Social Networks}.
\newblock \bibinfo{journal}{\emph{{IEEE} Trans. Computers}}
  \bibinfo{volume}{64}, \bibinfo{number}{3} (\bibinfo{year}{2015}),
  \bibinfo{pages}{640--653}.
\newblock


\bibitem[White et~al\mbox{.}(2023)]%
        {abs-2303-07839}
\bibfield{author}{\bibinfo{person}{Jules White}, \bibinfo{person}{Sam Hays},
  \bibinfo{person}{Quchen Fu}, \bibinfo{person}{Jesse Spencer{-}Smith}, {and}
  \bibinfo{person}{Douglas~C. Schmidt}.} \bibinfo{year}{2023}\natexlab{}.
\newblock \showarticletitle{ChatGPT Prompt Patterns for Improving Code Quality,
  Refactoring, Requirements Elicitation, and Software Design}.
\newblock \bibinfo{journal}{\emph{arXiv Preprint}} (\bibinfo{year}{2023}).
\newblock
\urldef\tempurl%
\url{https://arxiv.org/abs/2303.07839}
\showURL{%
\tempurl}


\bibitem[Zeng et~al\mbox{.}(2023)]%
        {ZengLDWL0YXZXTM23}
\bibfield{author}{\bibinfo{person}{Aohan Zeng}, \bibinfo{person}{Xiao Liu},
  \bibinfo{person}{Zhengxiao Du}, \bibinfo{person}{Zihan Wang},
  \bibinfo{person}{Hanyu Lai}, \bibinfo{person}{Ming Ding},
  \bibinfo{person}{Zhuoyi Yang}, \bibinfo{person}{Yifan Xu},
  \bibinfo{person}{Wendi Zheng}, \bibinfo{person}{Xiao Xia},
  \bibinfo{person}{Weng~Lam Tam}, \bibinfo{person}{Zixuan Ma},
  \bibinfo{person}{Yufei Xue}, \bibinfo{person}{Jidong Zhai},
  \bibinfo{person}{Wenguang Chen}, \bibinfo{person}{Zhiyuan Liu},
  \bibinfo{person}{Peng Zhang}, \bibinfo{person}{Yuxiao Dong}, {and}
  \bibinfo{person}{Jie Tang}.} \bibinfo{year}{2023}\natexlab{}.
\newblock \showarticletitle{{GLM-130B:} An Open Bilingual Pre-trained Model}.
  In \bibinfo{booktitle}{\emph{{ICLR}}}.
\newblock
\urldef\tempurl%
\url{https://openreview.net/forum?id=-Aw0rrrPUF}
\showURL{%
\tempurl}


\bibitem[Zhang et~al\mbox{.}(2023)]%
        {ZhangWZ0LHL23}
\bibfield{author}{\bibinfo{person}{Jian Zhang}, \bibinfo{person}{Xu Wang},
  \bibinfo{person}{Hongyu Zhang}, \bibinfo{person}{Hailong Sun},
  \bibinfo{person}{Xudong Liu}, \bibinfo{person}{Chunming Hu}, {and}
  \bibinfo{person}{Yang Liu}.} \bibinfo{year}{2023}\natexlab{}.
\newblock \showarticletitle{Detecting Condition-Related Bugs with Control Flow
  Graph Neural Network}. In \bibinfo{booktitle}{\emph{{ISSTA}}}.
  \bibinfo{pages}{1370--1382}.
\newblock


\bibitem[Zhang et~al\mbox{.}(2013)]%
        {ZhangXWZXG13}
\bibfield{author}{\bibinfo{person}{Jun Zhang}, \bibinfo{person}{Yang Xiang},
  \bibinfo{person}{Yu Wang}, \bibinfo{person}{Wanlei Zhou},
  \bibinfo{person}{Yong Xiang}, {and} \bibinfo{person}{Yong Guan}.}
  \bibinfo{year}{2013}\natexlab{}.
\newblock \showarticletitle{Network Traffic Classification Using Correlation
  Information}.
\newblock \bibinfo{journal}{\emph{{IEEE} Trans. Parallel Distributed Syst.}}
  \bibinfo{volume}{24}, \bibinfo{number}{1} (\bibinfo{year}{2013}),
  \bibinfo{pages}{104--117}.
\newblock


\bibitem[Zhang et~al\mbox{.}(2022)]%
        {abs-2205-01068}
\bibfield{author}{\bibinfo{person}{Susan Zhang}, \bibinfo{person}{Stephen
  Roller}, \bibinfo{person}{Naman Goyal}, \bibinfo{person}{Mikel Artetxe},
  \bibinfo{person}{Moya Chen}, \bibinfo{person}{Shuohui Chen},
  \bibinfo{person}{Christopher Dewan}, \bibinfo{person}{Mona~T. Diab},
  \bibinfo{person}{Xian Li}, \bibinfo{person}{Xi~Victoria Lin},
  \bibinfo{person}{Todor Mihaylov}, \bibinfo{person}{Myle Ott},
  \bibinfo{person}{Sam Shleifer}, \bibinfo{person}{Kurt Shuster},
  \bibinfo{person}{Daniel Simig}, \bibinfo{person}{Punit~Singh Koura},
  \bibinfo{person}{Anjali Sridhar}, \bibinfo{person}{Tianlu Wang}, {and}
  \bibinfo{person}{Luke Zettlemoyer}.} \bibinfo{year}{2022}\natexlab{}.
\newblock \showarticletitle{{OPT:} Open Pre-trained Transformer Language
  Models}.
\newblock \bibinfo{journal}{\emph{arXiv Preprint}} (\bibinfo{year}{2022}).
\newblock
\urldef\tempurl%
\url{https://arxiv.org/abs/2205.01068}
\showURL{%
\tempurl}


\bibitem[Zhao et~al\mbox{.}(2023)]%
        {abs-2303-18223}
\bibfield{author}{\bibinfo{person}{Wayne~Xin Zhao}, \bibinfo{person}{Kun Zhou},
  \bibinfo{person}{Junyi Li}, \bibinfo{person}{Tianyi Tang},
  \bibinfo{person}{Xiaolei Wang}, \bibinfo{person}{Yupeng Hou},
  \bibinfo{person}{Yingqian Min}, \bibinfo{person}{Beichen Zhang},
  \bibinfo{person}{Junjie Zhang}, \bibinfo{person}{Zican Dong},
  \bibinfo{person}{Yifan Du}, \bibinfo{person}{Chen Yang},
  \bibinfo{person}{Yushuo Chen}, \bibinfo{person}{Zhipeng Chen},
  \bibinfo{person}{Jinhao Jiang}, \bibinfo{person}{Ruiyang Ren},
  \bibinfo{person}{Yifan Li}, \bibinfo{person}{Xinyu Tang},
  \bibinfo{person}{Zikang Liu}, \bibinfo{person}{Peiyu Liu},
  \bibinfo{person}{Jian{-}Yun Nie}, {and} \bibinfo{person}{Ji{-}Rong Wen}.}
  \bibinfo{year}{2023}\natexlab{}.
\newblock \showarticletitle{A Survey of Large Language Models}.
\newblock \bibinfo{journal}{\emph{arXiv Preprint}} (\bibinfo{year}{2023}).
\newblock
\urldef\tempurl%
\url{https://arxiv.org/abs/2303.18223}
\showURL{%
\tempurl}


\bibitem[Zhao et~al\mbox{.}(2021)]%
        {ZhaoWFK021}
\bibfield{author}{\bibinfo{person}{Zihao Zhao}, \bibinfo{person}{Eric Wallace},
  \bibinfo{person}{Shi Feng}, \bibinfo{person}{Dan Klein}, {and}
  \bibinfo{person}{Sameer Singh}.} \bibinfo{year}{2021}\natexlab{}.
\newblock \showarticletitle{Calibrate Before Use: Improving Few-shot
  Performance of Language Models}. In \bibinfo{booktitle}{\emph{{ICML}}},
  Vol.~\bibinfo{volume}{139}. \bibinfo{pages}{12697--12706}.
\newblock


\bibitem[Zhou et~al\mbox{.}(2022)]%
        {ZhouYLL22}
\bibfield{author}{\bibinfo{person}{Kaiyang Zhou}, \bibinfo{person}{Jingkang
  Yang}, \bibinfo{person}{Chen~Change Loy}, {and} \bibinfo{person}{Ziwei Liu}.}
  \bibinfo{year}{2022}\natexlab{}.
\newblock \showarticletitle{Learning to Prompt for Vision-Language Models}.
\newblock \bibinfo{journal}{\emph{Int. J. Comput. Vis.}} \bibinfo{volume}{130},
  \bibinfo{number}{9} (\bibinfo{year}{2022}), \bibinfo{pages}{2337--2348}.
\newblock


\bibitem[Zhou et~al\mbox{.}(2019)]%
        {ZhouLSD019}
\bibfield{author}{\bibinfo{person}{Yaqin Zhou}, \bibinfo{person}{Shangqing
  Liu}, \bibinfo{person}{Jing~Kai Siow}, \bibinfo{person}{Xiaoning Du}, {and}
  \bibinfo{person}{Yang Liu}.} \bibinfo{year}{2019}\natexlab{}.
\newblock \showarticletitle{Devign: Effective Vulnerability Identification by
  Learning Comprehensive Program Semantics via Graph Neural Networks}. In
  \bibinfo{booktitle}{\emph{NeurIPS}}. \bibinfo{pages}{10197--10207}.
\newblock


\bibitem[Zou et~al\mbox{.}(2021)]%
        {ZouWXLJ21}
\bibfield{author}{\bibinfo{person}{Deqing Zou}, \bibinfo{person}{Sujuan Wang},
  \bibinfo{person}{Shouhuai Xu}, \bibinfo{person}{Zhen Li}, {and}
  \bibinfo{person}{Hai Jin}.} \bibinfo{year}{2021}\natexlab{}.
\newblock
  \showarticletitle{{\textdollar}{\textbackslash}mu{\textdollar}{\(\mu\)}VulDeePecker:
  {A} Deep Learning-Based System for Multiclass Vulnerability Detection}.
\newblock \bibinfo{journal}{\emph{{IEEE} Trans. Dependable Secur. Comput.}}
  \bibinfo{volume}{18}, \bibinfo{number}{5} (\bibinfo{year}{2021}),
  \bibinfo{pages}{2224--2236}.
\newblock


\end{thebibliography}
\end{document}